\documentclass[aps, prl, notitlepage, twocolumn, superscriptaddress]{revtex4-2}
\usepackage{xspace}
\usepackage{graphicx}
\usepackage{dcolumn}
\usepackage[usenames]{color}
\usepackage{slashed}
\usepackage[utf8]{inputenc}
\usepackage{amsfonts}
\usepackage{amsthm}
\usepackage{amsmath}
\usepackage{amssymb}
\usepackage{mathrsfs}
\usepackage{hyperref}
\usepackage{booktabs}
\usepackage{diagbox}
\usepackage{latexsym}
\usepackage{color}
\usepackage{bm}
\usepackage{CJK}
\usepackage{braket}

\usepackage{float}
\makeatletter
\let\newfloat\newfloat@ltx
\makeatother
\usepackage{algorithm}
\usepackage{algorithmicx}
\usepackage{algpseudocode}

\usepackage{caption}
\usepackage{subcaption}

\newcommand{\printfnsymbol}[1]{%
  \textsuperscript{\@fnsymbol{#1}}%
}

\begin{document}
\title{Non-Hermitian Delocalization Induced by Residue Imaginary Velocity}
\author{Shi-Xin Hu}
\affiliation{International Center for Quantum Materials, School of Physics, Peking University, Beijing 100871, China}
\author{Yongxu Fu}
\email{yongxufu@zjnu.edu.cn}
\affiliation{International Center for Quantum Materials, School of Physics, Peking University, Beijing 100871, China}
\affiliation{Department of Physics, Zhejiang Normal University, Jinhua, 321004, China}
\author{Yi Zhang}
\email{frankzhangyi@pku.edu.cn}
\affiliation{International Center for Quantum Materials, School of Physics, Peking University, Beijing 100871, China}
\affiliation{Collaborative Innovation Center of Quantum Matter, Beijing 100871, China}

\date{\today}

\begin{abstract}
The dichotomy of localization versus delocalization is a historic topic central to quantum and condensed matter physics. We discover a new delocalization mechanism attributed to a residue imaginary (part of) velocity $\operatorname{Im}(v)$, feasible for ground states or low-temperature states of non-Hermitian quantum systems under periodic boundary conditions. In sharp contrast to conventional formalisms through extended wavefunctions, these target systems exhibit delocalization in collective physical properties such as correlation and entanglement (of the Fermi Seas) despite sometimes localized left and right single-particle eigenstates, as we demonstrate numerically and generalize to scenarios with finite temperatures and interaction. Interestingly, disorder contributing to $\operatorname{Im}(v)$ may also allow strong-disorder delocalization. Thus, the nontrivial physics of $\operatorname{Im}(v)$ significantly enriches our understanding of delocalization and harbors interesting experiments and practical applications.
\end{abstract}

\maketitle

\emph{Introduction}--- Localization is a crucial physical concept with vital consequences on physical properties like correlation, transport, spectrum, and entanglement \cite{mott_1968, imada1998, mottbook, lee2006mott, roy2019mott, macKinnon1983, lee1985rev, kramer1993, brouwer1997, evers2008, quhall, laughlin1981, thouless1983, niu198401, niu1985, hatsugai199301, hatsugai199302}. For example, the Anderson localization arises in the presence of adequate disorder \cite{anderson1958, anderson1979, macKinnon1983, lee1985rev, kramer1993, brouwer1997, evers2008} - an infinitesimal disorder in a one-dimensional (1D) Hermitian system leads to localization across all eigenstates and correlation functions \cite{brouwer1997, evers2008}. Further, localization may occur due to interaction, e.g., in Mott insulators \cite{mott_1937, mott_1949, mott_1968, imada1998, mottbook, lee2006mott, roy2019mott, pan2021}, or simply the Pauli exclusion principle, e.g., in band insulators \cite{tknn, haldane1988,kane200501, kane200502, quspinhallwell, fu2007, castro2009, kitaev2009, hassan2010, qi2011, chiu2016}, where the occupied electron Fermi Seas display localization, e.g., short-range correlations and an absence of transport, despite fully extended single-particle Bloch wave functions.

Recent studies on non-Hermitian systems, originating from open quantum systems \cite{open1, open2, open3, open4, open5, open6, open7, open8}, optical systems (non-unitary quantum walk) \cite{optical1, optical2, optical3, optical4, optical5, optical6}, and electric circuits \cite{circuit1, circuit2, circuit3, circuit4, circuit5, circuit6}, have significantly broadened our understanding of condensed matter physics \cite{bergholtzrev2021, ashida2020, gong2018, kawabataprx}. The non-Hermitian skin effect (NHSE), which supports an extensive number of single-particle eigenstates localized at the boundaries and characterized by generalized Brillouin zones through the non-Bloch band theory \cite{yao2018, yokomizo2019, yang2020}, reveals an additional localization formalism in 1D non-Hermitian systems under open boundary conditions (OBCs) \cite{yao2018, yokomizo2019, yang2020, zhang2020, origin2020} and generalizable to higher dimensions \cite{liu2019second, lee2019ho, okugawa2020, kawabatahigher, fu2021, st2022, yokomizo2023nonbloch, wang2024amoeba, hu2023nonhermitianbandtheorydimensions, xiong2024nonhermitianskineffectarbitrary}. With the introduction of disorder in non-Hermitian systems, e.g., the 1D Hatano-Nelson (HN) model \cite{hatano1997}, a mobility edge coinciding with the parity-time (PT) transition may emerge yet eventually give way to complete Anderson localization in the strong-disorder limit \cite{Hatano1998, longhi2019prl, hui2019, longhi2019, liu2020general, liu2020pt, liu2021, yuce2022}. Nonetheless, such Anderson localization and NHSE physics are mainly limited to a single-particle framework and characterized by individual eigenstates and wave functions \cite{zeng2020, longhi202101, longhi202102, claes2021, okuma2021disorder, ronika2022, lin2022obser, orito2022en, kokkinakis2024anderson}, with only a few works focusing on non-Hermitian many-body systems \cite{PhysRevLett.123.090603, hamazaki2022lindbladianmanybodylocalization, PhysRevB.109.L140201, PhysRevB.109.174205}.

Here, we introduce a novel delocalization mechanism induced by a residue imaginary velocity $\operatorname{Im}(v)$. While the velocity expectation values for ground states or low-temperature states $\hat\rho\propto\exp\left(-\beta\hat H\right)$ generally possess a vanishing real part \footnote{The case of complex $\mu$ can be simplified to a real $\mu$ via a complex phase factor $e^{i\theta}\hat{H}$ upon the Hamiltonian and thus a rotation of the complex energy plane, as we elaborate in the Supplemental Material \cite{supp}. }, they may sustain a finite imaginary part $\operatorname{Im}(v)$ in non-Hermitian quantum systems under periodic boundary conditions (PBCs). Such a residue $\operatorname{Im}(v)$ mandates cumulative displacements in imaginary time, i.e., path integral, resulting in dominant worldlines traversing and winding around the system, as observed in quantum Monte Carlo stochastic series expansion (QMC-SSE) calculations \cite{PhysRevB.108.245114}. Consequently, a finite $\operatorname{Im}(v)$ offers straightforward criteria for delocalization, with corresponding physical behaviors such as power-law correlation and quasi-long-range entanglement.

Interestingly, there are models of disorders that may positively contribute to $\operatorname{Im}(v)$ and thus support delocalization. Consequently, such disorders' delocalization may prevail over its Anderson localization even in the strong-disorder limit, which was never possible previously. In contrast to conventional delocalized systems with extended wave functions bearing distant communications, we find scenarios with relatively strong disorder yet nonzero $\operatorname{Im}(v)$, where both the left and right single-particle eigenstates remain localized, and physical properties such as correlation and entanglement still exhibit delocalization. We reveal that such delocalization is contributed not by individual wave functions but collectively by the entire Fermi Sea. Therefore, the physics of $\operatorname{Im}(v)$, which is also straightforwardly generalizable to finite temperatures and interactions, greatly enriches our understanding of delocalization and bears potential experiments and applications, as we demonstrate in numerical examples.

\emph{Physics of the residue imaginary velocity}--- The expectation value of the velocity operator $\hat v = -i[\hat{x}, \hat H]$ for the ground state or low-temperature density operator $\hat{\rho}\propto \exp \left(-\beta \hat{H}\right)$ represents the macroscopic net velocity at equilibrium, which naturally vanishes for a Hermitian system $\hat H$. Non-Hermitian Hamiltonians, on the other hand, may effectively describe open or non-equilibrium (e.g., dissipative) quantum systems. Consequently, their $\hat{v}$ expectation values may yield nonzero net contributions under PBCs and bring about interesting physics \footnote{Admittedly, the microscopic mechanism that an open or non-equilibrium system may induce residue $\operatorname{Im}(v)$ like a non-Hermitian Hamiltonian remains an open question for future research. }. Following the Heisenberg equation $\hat{v}=d \hat{x}/d t$, our choice of the velocity operator is motivated by our ability to establish a direct connection between the single-particle spectrum, the velocity expectation value and the worldline winding, as well as consistency between Hermitian and non-Hermitian counterparts, as detailed in the following sections and the Supplementary Material \cite{supp}.

\begin{figure}
    \centering
    \includegraphics[scale=0.57]{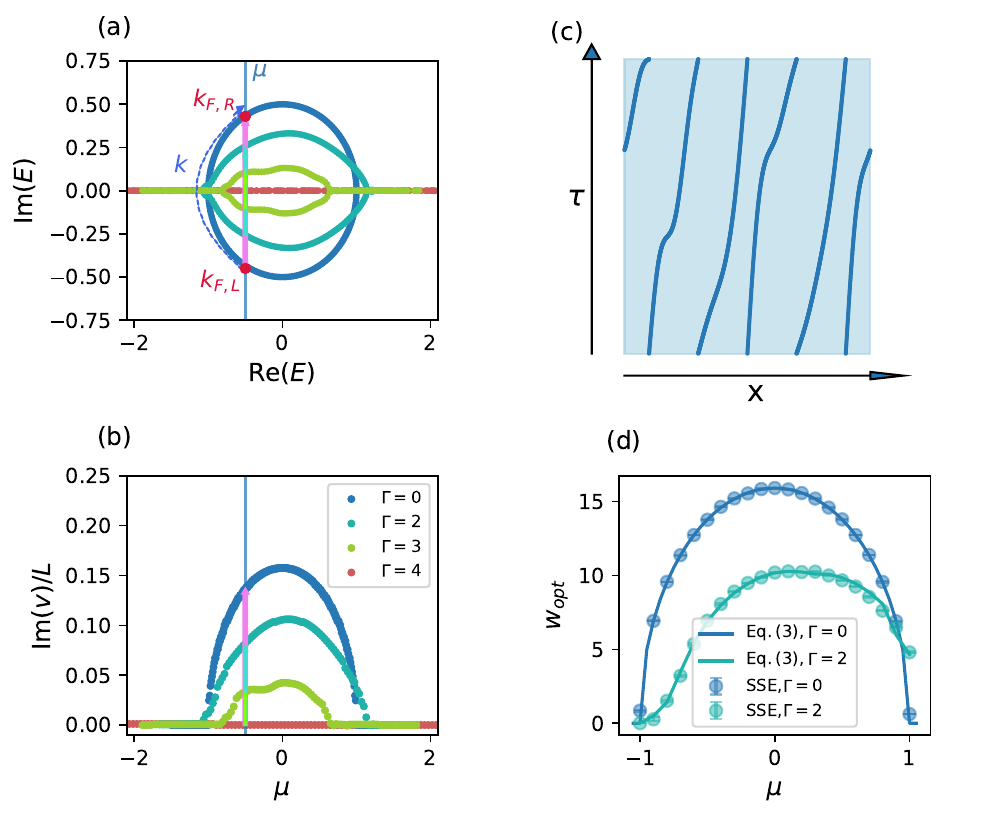}
    \caption{For the HN models under PBCs and different disorder strengths $\Gamma$, (a) the complex spectra and (b) the imaginary (part of) velocity $\operatorname{Im}(v)$ show general correspondences: the displacements (colored arrows) between Fermi points in the complex energy space dictate the residue $\operatorname{Im}(v)$ at that specific $\mu$. The dashed arrow indicates the direction of increasing $k$ for the pristine HN model. (c) The path-integral trajectories in $(1+1)$D space-time may circle the system, e.g., winding number $w_{opt}=1$, and (d) $w_{opt}$ of the HN models compare consistently between QMC-SSE calculations ($\beta=100$) and Eq. \ref{eq:woptvImv} with the corresponding $\operatorname{Im}(v)$. We set $\delta=0.5$ and $L=62$. }
    \label{fig:semiclassical}
\end{figure}

Let us consider the 1D HN model under PBC:
\begin{equation}
    \hat H_{\text{HN}}=\sum_x(1+\delta)c^{\dagger}_{x+1}c_x+(1-\delta)c^{\dagger}_x c_{x+1} = \sum_k  \epsilon_k c^{\dagger}_k c_k,
\end{equation}
where $c^\dagger_x$ and $c^\dagger_k$ are the fermion creation operator on site $x$ and its Fourier transform with momentum $k$, respectively. The PBC spectrum is $\epsilon_k= 2\cos k - 2i\delta \sin k$, as the deep-blue curve in Fig. \ref{fig:semiclassical}(a). For such a translation-invariant system, we can obtain its ground-state velocity:
\begin{equation}
v= \sum_{\operatorname{Re}(\epsilon_k)<\mu} v_{k} = \frac{L}{2\pi}  \left(\epsilon_{k_{F, R}} - \epsilon_{k_{F, L}}\right),
\end{equation}
where $v_{k}=\partial \epsilon_{k}/\partial k$. The summation over the Fermi Sea follows the real parts of the eigenenergies $\epsilon_k$, and $k_{F, R}$ ($k_{F, L}$) is the right (left) Fermi point \footnote{Multiple bands or Fermi Seas, if present, also need to be summed over. }; on the other hand, an alternative Fermi-Sea convention is the steady states after a long evolution time, which is dominated by $\operatorname{Im}$ - the imaginary parts of $\epsilon_k$; hereafter, we employ the former convention while noticing that these two Fermi-Sea conventions are interchangeable with an overall, $\pm i$ phase in front of the non-Hermitian Hamiltonian.

For a Hermitian system, e.g., the HN model with $\delta=0$, $\epsilon_{k} \in \mathbb{R}$ is real, and we have $\epsilon_{k_{F, R}} =\epsilon_{k_{F, L}} = \mu$, $v=0$. For a non-Hermitian system, on the other hand, $\epsilon_k \in \mathbb{C}$ is complex; while $\mu$ equalizes the Fermi energies' real parts $\operatorname{Re}(\epsilon_{k_{F, R}}) =\operatorname{Re}(\epsilon_{k_{F, L}}) = \mu$, their imaginary parts may differ $\operatorname{Im}(\epsilon_{k_{F, R}}) \neq \operatorname{Im}(\epsilon_{k_{F, L}})$, leading to a residue imaginary velocity $\operatorname{Im}(v)$ \footnote{It also implies that a residue $\operatorname{Im}(v)$ under PBC is related to the NHSE under OBC. }. Indeed, as $\mu$ varies, $\operatorname{Im}(v)$ consistently follows the corresponding traverse in the complex spectrum, and $\operatorname{Re}(v)$ always vanishes \cite{supp}; see Fig. \ref{fig:semiclassical}(b). Here, we have focused on the ground state $\beta\rightarrow \infty$, but the arguments also apply to low-temperature scenarios, as we will discuss later.

Such a residue $\operatorname{Im}(v)$ has far-reaching physical impacts. Semiclassically, its quantum dynamics characterizes a unidirectional displacement in imaginary time $\tau = it \in \mathbb{R}$, i.e., relevant worldlines in $(1+1)$D path integral incline to shift cumulatively and circle the system [Fig. \ref{fig:semiclassical}(c)]:
\begin{equation}
w_{opt} = \frac{dx}{d\tau} \cdot \frac{\beta}{L} = \operatorname{Im}(v) \cdot \beta / L,
\label{eq:woptvImv}
\end{equation}
where $w_{opt}$ is the winding number upon an imaginary time $\beta$, and $L$ is the system size. Indeed, Fig. \ref{fig:semiclassical}(d) shows such nontrivial $w_{opt}$, obtained in QMC-SSE calculations of the HN models under PBCs \cite{PhysRevB.108.245114}, compare consistently with $\operatorname{Im}(v)$. Importantly, such $\operatorname{Im}(v)\neq 0$ mandates dominant path-integral trajectories traverse the entire system, causing delocalization that is reflected in physical behaviors, e.g., correlation and entanglement, as we will show later and in the Supplemental Material \cite{supp}. On the other hand, such a mechanism also requires PBCs, as charge conservation in an equilibrium system under OBC will enforce a vanishing stable current $v=0$ and a zero winding number $w_{opt}=0$.

The physics of $\operatorname{Im}(v)$ also applies to disordered systems without translation symmetry, where we can substitute the single-particle momentum $k$ with the phase $\phi$ across the periodic boundary. We have included results of the HN models with random disorder $\hat H_{dis} = \sum_x \gamma_x c^\dagger_x c_x$, $\gamma_x \in [-\Gamma, \Gamma]$, in Fig. \ref{fig:semiclassical}. The Anderson localization competes with $\operatorname{Im}(v)$; as $\Gamma$ increases and gradually overwhelms the delocalization effect, the complex spectrum loop shrinks, and accordingly, $\operatorname{Im}(v)$'s amplitude and allowed $\mu$ window decreases \footnote{Similarly to Hermitian quantum systems, localization begins at the band edges and moves toward the center. }, until the entire spectrum collapses into a line and $\operatorname{Im}(v)$ vanishes globally at $\Gamma\geq4$. We note that $\operatorname{Im}(v)$ serves as both a mechanism for and a straightforward signature of delocalization.

\begin{figure}
    \centering
    \includegraphics[scale=0.53]{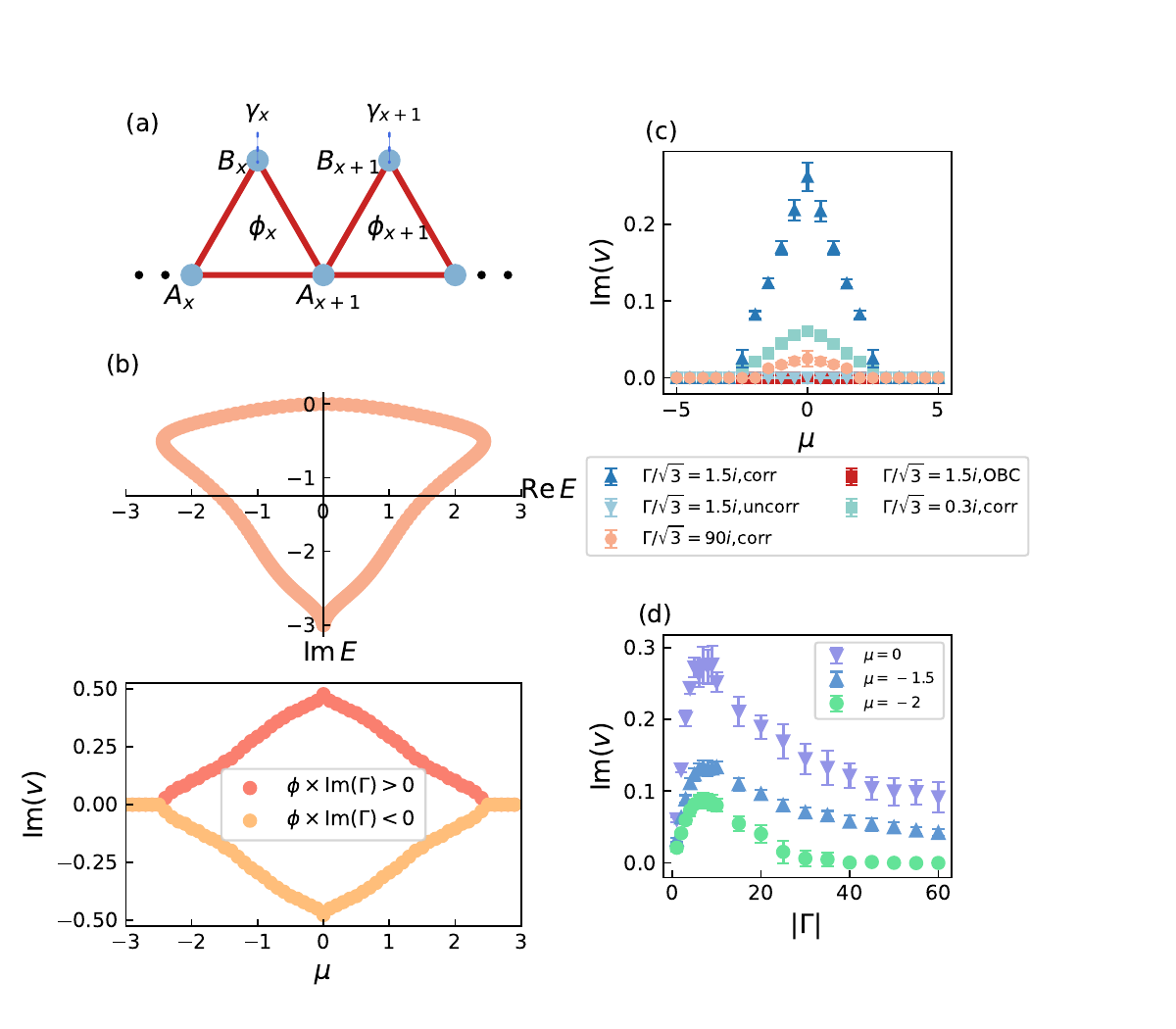}
    \caption{(a) The model in Eq. \ref{eq:ham} consists of an on-site potential $\gamma(x)$ on the $B$ sublattice and a flux $\phi(x)$ though the triangle on each lattice site $x$. (b) The complex spectrum and $\operatorname{Im}(v)$ of a translation invariant model with $\gamma(x)=3i$ and $\phi(x)=\pi/2$. $\operatorname{Im}(v)$ changes sign as we flip $\gamma(x)=\pm3i$ or $\phi(x)=\pm\pi/2$. (c) and (d): While $\operatorname{Im}(v)\neq 0$ within a wide range of $\mu$ for various imaginary $\Gamma$ and correlated phase $\phi(x)$ under PBCs, even strong disorder $|\Gamma|\gg 1$, it vanishes for uncorrelated $\phi(x)$ or OBCs. }
    \label{fig:model}
\end{figure}

\emph{Delocalization behaviors and strong-disorder limit}--- Inspired by the dissipative fluxed model in Ref. \onlinecite{triangleNHmodel}, we consider a 1D non-Hermitian Hamiltonian as follows:
\begin{equation}
\hat{H}_{AB} = \sum_x (a_{x+1}^\dagger a_{x} + b_x^\dagger a_x + a_{x+1}^\dagger b_x e^{i\phi(x)}  +\operatorname{h.c.}) + \gamma(x) b_x^\dagger b_x,
\label{eq:ham}
\end{equation}
where $a^\dagger_x$ ($b^\dagger_x$) is the fermion creation operator on the A (B) sublattice of site $x$, and $\gamma(x)$ [$\phi(x)$] is an onsite potential (flux); see illustration in Fig. \ref{fig:model}(a). Fig. \ref{fig:model}(b) shows its complex spectrum and residue imaginary velocity $\operatorname{Im}(v)$ with translation-invariant $\gamma(x)=3i$, $\phi(x)=\pi/2$ and PBC. As we invert either $\gamma(x)$ or $\phi(x)$, equivalent to a Hermitian or inversion transformation, respectively, $\operatorname{Im}(v)$ changes sign. In contrast, a real $\gamma(x)$ thus Hermitian $\hat{H}_{AB}$ does not impose a finite $\operatorname{Im}(v)$. We discuss more general settings of $\hat{H}_{AB}$ in the Supplemental Material \cite{supp}. Once again, we emphasize that a finite $\operatorname{Im}(v)$ is only possible under PBCs, as both the velocity $v$ and the path-integral winding number $w_{opt}$ are bound to vanish under OBCs, even if the rest of the settings are identical - in sharp contrast with previous delocalization generally insensitive to boundary conditions.

Next, we introduce randomness in $\gamma(x)=\Gamma U(x)$ with $U(x)\in\left[-1, 1\right]$, which naturally gives rise to the Anderson localization and thus a vanishing $\operatorname{Im}(v)$. On the other hand, if we align the signs of $\phi(x)$ with $\gamma(x)$:
\begin{equation}
    \phi(x)=\left\{
\begin{aligned}
\pi/2 & , & U(x)>0, \\
-\pi/2 & , & U(x)<0,
\end{aligned}
\right. \label{eq:corphi}
\end{equation}
so that such correlated disorder keeps consistent contributions to $\operatorname{Im}(v)$, and its delocalization effect may prevail over the Anderson localization. Indeed, $\operatorname{Im}(v)$ remains finite within a range of Fermi energy $\mu$, given imaginary $\Gamma$, correlated $\phi(x)$, and PBCs; see Figs. \ref{fig:model}(c) and \ref{fig:model}(d) for a summary of the results. Interestingly, we obtain a strong-disorder-limit delocalization for the first time: $\operatorname{Im}(v)$ survives at strong disorder $|\Gamma|\gg 1$ [Fig. \ref{fig:model}(d)]; see a semi-quantitative analysis of $\operatorname{Im}(v)$ in the strong-disorder limit in the Supplemental Material \cite{supp}. Such a delocalized behavior is in sheer contrast with the HN model, where the disorder contributes only to the Anderson localization and suppresses whatever non-disorder-based delocalization in the strong-disorder limit \cite{hui2019, longhi2019, longhi2019prl, liu2020general, liu2020pt, liu2021, yuce2022}.

\begin{figure}
    \centering
    \includegraphics[scale=0.5]{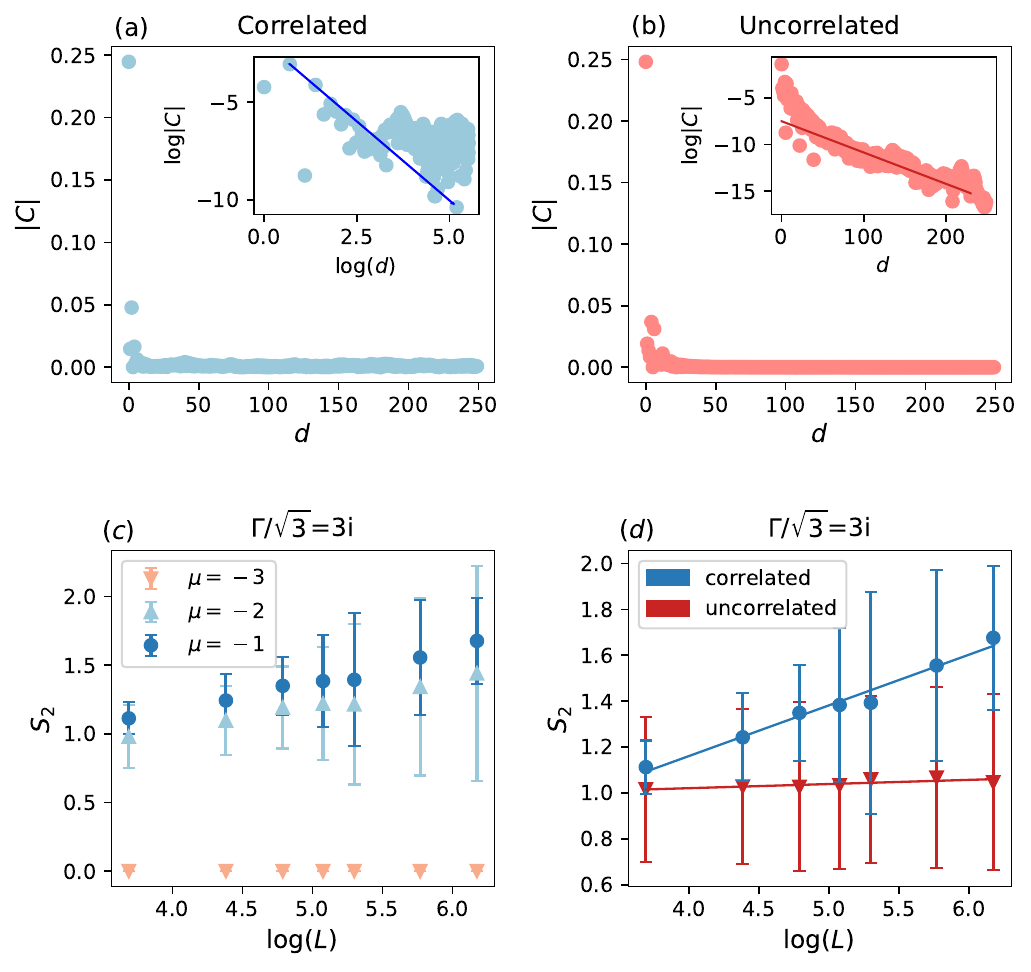}
    \caption{The correlation functions between the $A$ sublattices of $\hat{H}_{AB}$ in Eq. \ref{eq:ham} exhibit (a) a power-law scaling for phase-correlated disorder (Eq. \ref{eq:corphi}) and (b) an exponential decay for uncorrelated disorder. We set $\Gamma = 1.5\times\sqrt{3}i$, $\mu=-1$, and the system size $L=1000$. The insets are log-log and log-linear plots, respectively. Likewise, the second Renyi entropy $S_2$ shows an area law for the localized scenarios [$\mu=-3$ case in (c) and uncorrelated case in (d)] and a logarithmic correction, i.e., $S_2\propto \log(L)$, for the delocalized scenarios [$\mu=-1, -2$ cases in (c) and the phase-correlated case in (d)]. We implement phase-correlated disorder in (c) and $\mu=-1$ in (d). }
    \label{fig:corr_pbc_N1000}
\end{figure}

The delocalization exhibits clear physical signatures in correlation and entanglement: the correlation functions $C(d)=\frac{1}{N}\sum_x \langle c^{\dagger}_x c_{x+d}\rangle$ exhibit a power-law scaling with respect to spatial distance $d$ in the delocalized case where $\operatorname{Im}(v)$ remains finite [Fig. \ref{fig:corr_pbc_N1000}(a)] yet an exponential decay in the localized case where $\operatorname{Im}(v)$ vanishes [Fig. \ref{fig:corr_pbc_N1000}(b)]; the second Renyi entropy $S_2=-\operatorname{log} \operatorname{tr}(\hat \rho^2)$ also exhibit quasi-long-range behaviors with logarithmic corrections $S_2 \propto \operatorname{log}(L)$ [$\mu=-1$ and $\mu=-2$ in Fig. \ref{fig:corr_pbc_N1000}(c), correlated case in Fig. \ref{fig:corr_pbc_N1000}(d)] in scenarios that coincide with a finite $\operatorname{Im}(v)$, instead of Area-Law behaviors otherwise [$\mu=-3$ in Fig. \ref{fig:corr_pbc_N1000}(c), uncorrelated case in Fig. \ref{fig:corr_pbc_N1000}(d)]. Here, $\hat \rho$ is the reduced density operator on half of the system \cite{PhysRevB.108.245114}. We note that compared with $C(d)$ or $S_2$, $v$ concerns mostly local operators and is thus much more straightforward to evaluate for identifying (de)localization.

\emph{``Fermi-sea" delocalization}--- A common defining feature of delocalization is the extended wave functions. Such a criterion is also generalizable to left and right eigenstates in non-Hermitian systems, e.g., single-particle states become exponentially localized around respective sites once the Anderson localization dominates over the NHSE under OBCs and strong disorder \cite{hui2019, longhi2019, liu2020general, liu2020pt, zeng2020, longhi202101, longhi202102, liu2021, yuce2022}. It is believed that eigenstates are more likely delocalized with finite $\operatorname{Im}(E)$ on the spectrum loops under PBCs \cite{Hatano1998, longhi2019prl, liu2020general, liu2020pt}; however, a finite $\operatorname{Im}(E)$ does not guarantee delocalized wave functions \cite{Ma2025}, and despite localized left $\psi_{nL}(x)$ and right $\psi_{nL}(x)$ eigenstates, their inner product $\psi^{*}_{nL}(x) \psi_{nR}(x)$ may still exhibit certain delocalized behaviors \cite{Hatano1998}.

Indeed, the single-particle wave functions of non-Hermitian systems $\hat H_{AB}$ and $\hat H_{HN}$ in weak disorder exhibit extended spatial distributions, delocalized physical properties, together with a residue $\operatorname{Im}(v)$. On the other hand, the single-particle left $\psi_{nL}(x)$ and right eigenstates $\psi_{nR}(x)$ behave localized in strong disorder, despite finite $\operatorname{Im}(v)$ and (quasi-)long-range correlations and entanglement; see Fig. \ref{fig:Eigen_IPR}(a), Table \ref{table:locase}, and the inverse participation ratio (IPR) analysis in the Supplemental Material \cite{supp}. The delocalization of $\operatorname{Im}(v)$ and physical properties of non-Hermitian systems in strong disorder may follow a fully distinctive mechanism from extended single-particle wave functions.

\begin{figure}
    \centering
    \includegraphics[scale=0.65]{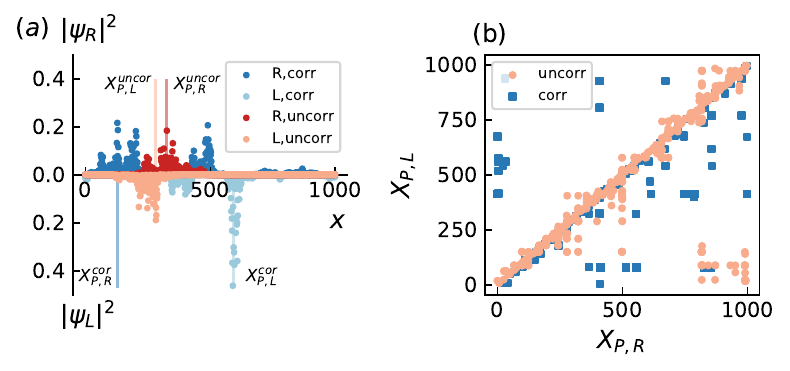}
    \caption{(a) Typical distributions of a pair of left and right eigenstates of $\hat{H}_{AB}$ with strong disorder $\Gamma=5i$ show localized single-particle wave functions irrespective of correlated or uncorrelated phase $\phi(x)$, thus a residue $\operatorname{Im}(v)$ or not. Therefore, the former's delocalized physical properties trace back to a different origin: while left and right eigenstates shadow each other relatively closely in the latter case, they may separate globally in the former case, as indicated in (b) the correspondence between $X_{P, L}$ and $X_{P, R}$ - the peak locations of left and right eigenstates - for the former case ($\operatorname{Im}(v)\neq 0$), the latter case ($\operatorname{Im}(v)= 0$), and the Hermitian case $\psi_{nL}(x) = \psi^{*}_{nR}(x)$. Note that $X_{P, R} \sim X_{P, L}$ at the top left and bottom right corners due to PBCs. A corresponding, more quantitative $\psi^{*}_{nL}(x)\psi_{nR}(x')$ heat map (Fig. S8) is in Ref. \cite{supp}. }
    \label{fig:Eigen_IPR}
\end{figure}

\begin{table}
\centering
\begin{tabular}{ |c|c|c| }
 \hline
  & Weak disorder & Strong disorder \\ \hline
 $\operatorname{Im}(v) = 0$ & Localized wave functions & Localized wave functions \\
  & and localized properties & and localized properties   \\ \hline
 $\operatorname{Im}(v)\neq 0$ & Extended wave functions & Localized wave functions \\
  & and delocalized properties & \emph{yet} delocalized properties \\ \hline
\end{tabular}
\caption{We summarize the localizability of single-particle wave functions and physical properties in cases of $\operatorname{Im}(v)$ and disorder strength. The seeming contradiction for the strong-disorder $\operatorname{Im}(v)\neq 0$ case suggests an unconventional delocalization mechanism. }
\label{table:locase}
\end{table}

In strong disorder, both the left $\psi_{nL}(x)$ and right $\psi_{nR}(x)$ eigenstates are localized, limiting the extension of the outer product, $\psi^{*}_{nL}(x)\psi_{nR}(x')$. In Hermitian systems, $\psi_{nL}(x)=\psi^*_{nR}(x)$; non-Hermitian systems allow $\psi_{nL}(x)$ and $\psi_{nR}(x)$ to differ, yet they generally shadow each other closely and differ only locally when $\operatorname{Im}(v)$ vanishes; in contrast, when delocalization looms given $\operatorname{Im}(v)\neq 0$, a pair of $\psi_{nL}(x)$ and right $\psi_{nR}(x)$ may reside in globally different and vastly distant regions, as are apparent from their respective peak positions $X_{P, R}$ and $X_{P, L}$ in Fig. \ref{fig:Eigen_IPR}(b) and the heat map of $\psi^{*}_{nL}(x)\psi_{nR}(x')$ in the Supplemental Material \cite{supp}. While an eigenstate merely communicates a pair of remote spots, another eigenstate connects another pair of spots, and so on, until global contacts and delocalizations are eventually achieved by collective contributions of numerous eigenstates across the occupied Fermi Sea. This delocalization thus appears in physical properties related to the Fermi Sea, such as the density operator $\hat \rho \propto \exp(-\beta \hat H)$, the path-integral worldlines, correlations, and entanglement. Its advent is likely a crossover, as we observe the gradual localization of the wave functions as the disorder strength increases or the gradual increase of $|X_{P, R}-X_{P, L}|$ displacement as $\operatorname{Im}(v)$ kicks in; see detailed results (Figs. S7 - S9) in the Supplemental Material \cite{supp}.

\begin{figure}
    \centering
    \includegraphics[scale=0.65]{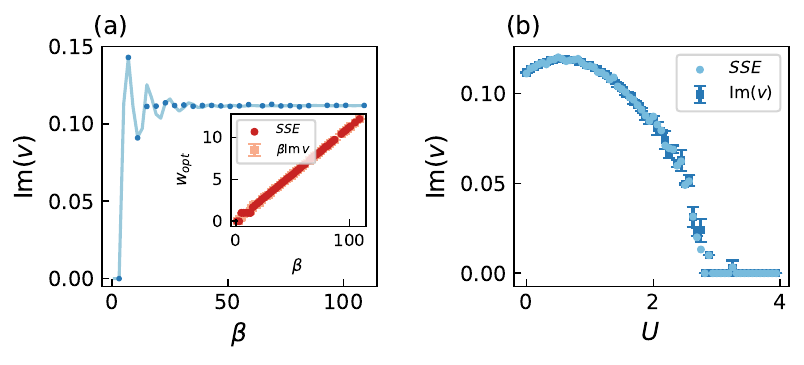}
    \caption{(a) The residue imaginary velocity $\operatorname{Im}(v)$ of the HN model $\hat H_{HN}$ initiates from 0 at small $\beta$ (high temperature) and eventually converges to a finite value at large $\beta$ (low temperature). (b) The repulsive interaction $U$ suppresses the delocalization and fully eliminate the residue $\operatorname{Im}(v)$ at $U>3$. $\operatorname{Im}(v)$ trends consistently with the winding number $w_{opt}$ of path-integral worldlines (dots) and the correlation behaviors \cite{supp} in both scenarios. We set Fermi energy $\mu=0$, non-Hermitian $\delta=0.5$, and disorder $\Gamma=2$. }
    \label{fig:Imv_varbeta}
\end{figure}

Such a Fermi-sea nature, irrespective of single-particle wave functions' extensiveness \footnote{On the contrary, localized physical properties require both localized single-particle states and $\operatorname{Im}(v) = 0$. }, also suggests straightforward generalizations to finite temperature and interaction. At finite $\beta=1/k_B T$, $\hat \rho = \sum_n f(\epsilon_n)|nR\rangle \langle nL|$, where $f(\epsilon_n) =  1/(\exp^{\beta \epsilon_n}+1)$ tends to 1 (0) for $\operatorname{Re}(\epsilon_n) \ll k_B T$ [$\operatorname{Re}(\epsilon_n) \gg -k_B T$] regardless of the imaginary part, filling the Fermi Sea. At low $T$, only few states within the narrow energy window $|\operatorname{Re}(\epsilon_n)| \lesssim k_B T$ deviates from the $T=0$ case - a residue $\operatorname{Im}(v)$ and corresponding delocalization will still emerge; on the contrary, at high temperatures and small $\beta$, the path-integral worldlines may not possess sufficient imaginary time span ($\beta$) to get across the system; thus, we expect an impaired $\operatorname{Im}(v)$ delocalization. Indeed, the results of $\operatorname{Im}(v)$ for the HN model at various $\beta$ in Fig. \ref{fig:Imv_varbeta}(a) verify our expectations. We also consider the HN model with the interaction $\hat H_U=U\sum_x \hat n_x \hat n_{x+1}$, where $\hat n_x= c^\dagger_x c_x$. A repulsive interaction $U>0$ competes with $\operatorname{Im}(v)$ and gradually drives the quantum many-body system toward localization in a Mott-insulating fashion, as is summarized in Fig. \ref{fig:Imv_varbeta}(b) and Ref. \cite{supp}.

\emph{Conclusions and discussions}--- We have discussed a new delocalization mechanism and signature attributed to a residue imaginary velocity $\operatorname{Im}(v)$ for non-Hermitian quantum systems under PBCs. We have demonstrated its presence, together with characteristic correlation and entanglement, in various models, including finite-temperature and interacting scenarios. Such $\operatorname{Im}(v)$ may exceed the Anderson localization and sustain delocalization in strong disorder, even when the corresponding single-particle eigenstates are localized. The nontrivial physics of $\operatorname{Im}(v)$ significantly enriches our existing knowledge of delocalization.

Such $\operatorname{Im}(v)$ delocalization may also apply to real-time evolution: despite localized wave functions of certain non-Hermitian $\hat H$ under strong disorder, $e^{-i\hat{H}t}=e^{-i\epsilon_n t}| \psi^R_n \rangle \langle\psi^L_n |$ behaves non-local. In addition, there are interesting applications, e.g., quantum simulations and optimizations. For instance, a quantum adiabatic process obtains a target state by interpolating the corresponding Hamiltonian $\hat H_1$ with an initial $\hat H_0$: $\hat H_{\lambda} = \lambda \hat H_1 + (1-\lambda) \hat H_0$, $\lambda \in [0, 1]$. In practice, however, such formalism breaks down when the ground state changes abruptly, commonly due to localized states around distantly separated minima. A residue $\operatorname{Im}(v)$ may bring forth delocalization and, thus, smoother evolution.

\begin{figure}
    \centering
    \includegraphics[scale=0.6]{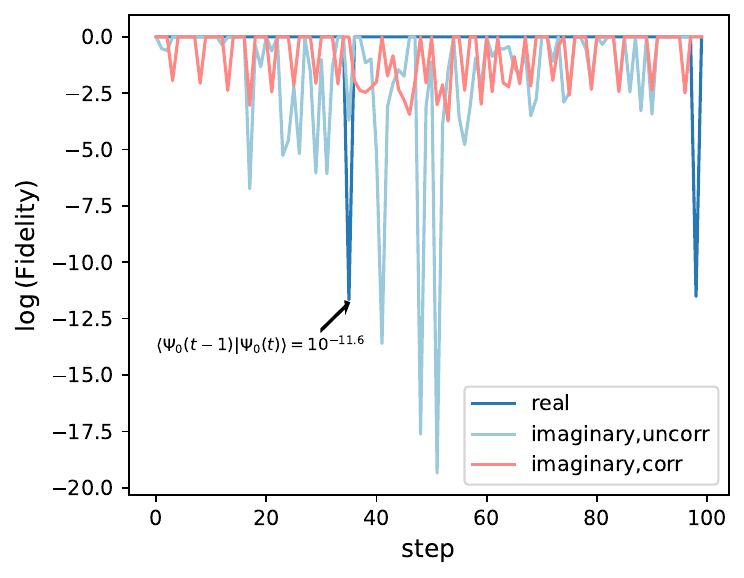}
    \caption{The fidelity along paths interpolating between disordered Hamiltonians indicates significantly better adiabaticity in the presence of $\operatorname{Im}(v)$ delocalization: unlike the localized cases [uncorrelated phase $\phi(x)$ or real $\Gamma$] where the ground states suffer abrupt changes at isolated instances, the delocalized case [$\Gamma$ with imaginary part and correlated $\phi(x)$] may distribute the evolution more evenly along the path. }
    \label{fig:qaa}
\end{figure}

For example, we interpolate disordered models in Eq. \ref{eq:ham} under PBCs and evaluate the fidelity between ground states along the paths. As we summarize in Fig. \ref{fig:qaa}, when and only when we successfully establish a residue $\operatorname{Im}(v)$, i.e., $\Gamma$ with an imaginary part and correlated phase $\phi(x)$, the ground states (i.e., Fermi seas) evolve relatively smoothly and evenly along the process, while the other cases commonly witness abrupt changes thus collapsed fidelity at one point or another; see more detailed settings and results in Supplemental Material \cite{supp}. Notably, such advantage is only present under PBCs but not OBCs.

We note that our work focuses on non-Hermitian properties at equilibrium or large-time-scale non-Hermitian dynamics, thus difficult from the short-time non-Hermitian effective theories for open and dissipative quantum systems \cite{ECHEVERRIARTEAGA2018413, Reisenbauer2024}; it is more compatible with the quantum adiabatic simulations or time evolution \cite{PhysRevB.108.214308, longhi202102} or with suppressed quantum jumped through quantum-circuit simulations \cite{PhysRevB.110.035113, ZhangYuxuan2025}. On the other hand, cold atom \cite{Boyan2020, Lapp2019}, optical \cite{Xue2024, optical4, optical5}, acoustic \cite{gu2021controlling, Zhang2021, Wen2022, fan2023reconf, Huang2024}, mechanical \cite{brandenbourger2019, ananya2020, chen2021, wang2022morphing, wang2023exp, li2024obser}, and electric systems \cite{circuit1, circuit2, circuit3, circuit4} can directly simulate non-Hermitian single-particle phenomena such as the NHSE \cite{PhysRevLett.125.187403, PhysRevResearch.6.L012061}; however, they are classical or bosonic systems, making Fermi-sea properties out of reach, and mostly without interactions. Nevertheless, the exotic physics of localized left and right single-particle eigenstates serving as consequences of $\operatorname{Im}(v)$, as in Fig. 4, are entirely accessible via these platforms once the PBC setups are adequately addressed in practice. Finally, the Fermi-sea or many-body physics of $\operatorname{Im}(v)$ may be realizable in novel solid-state systems, e.g., multi-layer graphene and TMD materials \cite{PhysRevB.107.035306, PhysRevLett.132.156301}.

\emph{Acknowledgment:} We acknowledge generous support from the National Key R\&D Program of China (Grant No.2022YFA1403700) and the National Natural Science Foundation of China (Grants No.12174008 \& No.92270102).

\bibliography{ref.bib}

\begin{thebibliography}{127}%
\makeatletter
\providecommand \@ifxundefined [1]{%
 \@ifx{#1\undefined}
}%
\providecommand \@ifnum [1]{%
 \ifnum #1\expandafter \@firstoftwo
 \else \expandafter \@secondoftwo
 \fi
}%
\providecommand \@ifx [1]{%
 \ifx #1\expandafter \@firstoftwo
 \else \expandafter \@secondoftwo
 \fi
}%
\providecommand \natexlab [1]{#1}%
\providecommand \enquote  [1]{``#1''}%
\providecommand \bibnamefont  [1]{#1}%
\providecommand \bibfnamefont [1]{#1}%
\providecommand \citenamefont [1]{#1}%
\providecommand \href@noop [0]{\@secondoftwo}%
\providecommand \href [0]{\begingroup \@sanitize@url \@href}%
\providecommand \@href[1]{\@@startlink{#1}\@@href}%
\providecommand \@@href[1]{\endgroup#1\@@endlink}%
\providecommand \@sanitize@url [0]{\catcode `\\12\catcode `\$12\catcode
  `\&12\catcode `\#12\catcode `\^12\catcode `\_12\catcode `\%12\relax}%
\providecommand \@@startlink[1]{}%
\providecommand \@@endlink[0]{}%
\providecommand \url  [0]{\begingroup\@sanitize@url \@url }%
\providecommand \@url [1]{\endgroup\@href {#1}{\urlprefix }}%
\providecommand \urlprefix  [0]{URL }%
\providecommand \Eprint [0]{\href }%
\providecommand \doibase [0]{https://doi.org/}%
\providecommand \selectlanguage [0]{\@gobble}%
\providecommand \bibinfo  [0]{\@secondoftwo}%
\providecommand \bibfield  [0]{\@secondoftwo}%
\providecommand \translation [1]{[#1]}%
\providecommand \BibitemOpen [0]{}%
\providecommand \bibitemStop [0]{}%
\providecommand \bibitemNoStop [0]{.\EOS\space}%
\providecommand \EOS [0]{\spacefactor3000\relax}%
\providecommand \BibitemShut  [1]{\csname bibitem#1\endcsname}%
\let\auto@bib@innerbib\@empty
\bibitem [{\citenamefont {MOTT}(1968)}]{mott_1968}%
  \BibitemOpen
  \bibfield  {author} {\bibinfo {author} {\bibfnamefont {N.~F.}\ \bibnamefont
  {MOTT}},\ }\bibfield  {title} {\bibinfo {title} {Metal-insulator
  transition},\ }\href {https://doi.org/10.1103/RevModPhys.40.677} {\bibfield
  {journal} {\bibinfo  {journal} {Rev. Mod. Phys.}\ }\textbf {\bibinfo {volume}
  {40}},\ \bibinfo {pages} {677} (\bibinfo {year} {1968})}\BibitemShut
  {NoStop}%
\bibitem [{\citenamefont {Imada}\ \emph {et~al.}(1998)\citenamefont {Imada},
  \citenamefont {Fujimori},\ and\ \citenamefont {Tokura}}]{imada1998}%
  \BibitemOpen
  \bibfield  {author} {\bibinfo {author} {\bibfnamefont {M.}~\bibnamefont
  {Imada}}, \bibinfo {author} {\bibfnamefont {A.}~\bibnamefont {Fujimori}},\
  and\ \bibinfo {author} {\bibfnamefont {Y.}~\bibnamefont {Tokura}},\
  }\bibfield  {title} {\bibinfo {title} {Metal-insulator transitions},\ }\href
  {https://doi.org/10.1103/RevModPhys.70.1039} {\bibfield  {journal} {\bibinfo
  {journal} {Rev. Mod. Phys.}\ }\textbf {\bibinfo {volume} {70}},\ \bibinfo
  {pages} {1039} (\bibinfo {year} {1998})}\BibitemShut {NoStop}%
\bibitem [{\citenamefont {Fazekas}(1999)}]{mottbook}%
  \BibitemOpen
  \bibfield  {author} {\bibinfo {author} {\bibfnamefont {P.}~\bibnamefont
  {Fazekas}},\ }\href {https://doi.org/10.1142/2945} {\emph {\bibinfo {title}
  {Lecture Notes on Electron Correlation and Magnetism}}}\ (\bibinfo
  {publisher} {World Scientific},\ \bibinfo {address} {Singapore},\ \bibinfo
  {year} {1999})\BibitemShut {NoStop}%
\bibitem [{\citenamefont {Lee}\ \emph {et~al.}(2006)\citenamefont {Lee},
  \citenamefont {Nagaosa},\ and\ \citenamefont {Wen}}]{lee2006mott}%
  \BibitemOpen
  \bibfield  {author} {\bibinfo {author} {\bibfnamefont {P.~A.}\ \bibnamefont
  {Lee}}, \bibinfo {author} {\bibfnamefont {N.}~\bibnamefont {Nagaosa}},\ and\
  \bibinfo {author} {\bibfnamefont {X.-G.}\ \bibnamefont {Wen}},\ }\bibfield
  {title} {\bibinfo {title} {Doping a mott insulator: Physics of
  high-temperature superconductivity},\ }\href
  {https://doi.org/10.1103/RevModPhys.78.17} {\bibfield  {journal} {\bibinfo
  {journal} {Rev. Mod. Phys.}\ }\textbf {\bibinfo {volume} {78}},\ \bibinfo
  {pages} {17} (\bibinfo {year} {2006})}\BibitemShut {NoStop}%
\bibitem [{\citenamefont {Roy}(2019)}]{roy2019mott}%
  \BibitemOpen
  \bibfield  {author} {\bibinfo {author} {\bibfnamefont {S.~B.}\ \bibnamefont
  {Roy}},\ }\bibfield  {title} {\bibinfo {title} {Mott insulators and related
  phenomena: a basic introduction},\ }in\ \href
  {https://doi.org/10.1088/2053-2563/ab16c9ch3} {\emph {\bibinfo {booktitle}
  {Mott Insulators}}},\ \bibinfo {series and number} {2053-2563}\ (\bibinfo
  {publisher} {IOP Publishing},\ \bibinfo {year} {2019})\ pp.\ \bibinfo {pages}
  {3--1 to 3--35}\BibitemShut {NoStop}%
\bibitem [{\citenamefont {MacKinnon}\ and\ \citenamefont
  {Kramer}(1983)}]{macKinnon1983}%
  \BibitemOpen
  \bibfield  {author} {\bibinfo {author} {\bibfnamefont {A.}~\bibnamefont
  {MacKinnon}}\ and\ \bibinfo {author} {\bibfnamefont {B.}~\bibnamefont
  {Kramer}},\ }\bibfield  {title} {\bibinfo {title} {The scaling theory of
  electrons in disordered solids: Additional numerical results},\ }\href
  {https://doi.org/10.1007/BF01578242} {\bibfield  {journal} {\bibinfo
  {journal} {Zeitschrift f{\"u}r Physik B Condensed Matter}\ }\textbf {\bibinfo
  {volume} {53}},\ \bibinfo {pages} {1} (\bibinfo {year} {1983})}\BibitemShut
  {NoStop}%
\bibitem [{\citenamefont {Lee}\ and\ \citenamefont
  {Ramakrishnan}(1985)}]{lee1985rev}%
  \BibitemOpen
  \bibfield  {author} {\bibinfo {author} {\bibfnamefont {P.~A.}\ \bibnamefont
  {Lee}}\ and\ \bibinfo {author} {\bibfnamefont {T.~V.}\ \bibnamefont
  {Ramakrishnan}},\ }\bibfield  {title} {\bibinfo {title} {Disordered
  electronic systems},\ }\href {https://doi.org/10.1103/RevModPhys.57.287}
  {\bibfield  {journal} {\bibinfo  {journal} {Rev. Mod. Phys.}\ }\textbf
  {\bibinfo {volume} {57}},\ \bibinfo {pages} {287} (\bibinfo {year}
  {1985})}\BibitemShut {NoStop}%
\bibitem [{\citenamefont {Kramer}\ and\ \citenamefont
  {MacKinnon}(1993)}]{kramer1993}%
  \BibitemOpen
  \bibfield  {author} {\bibinfo {author} {\bibfnamefont {B.}~\bibnamefont
  {Kramer}}\ and\ \bibinfo {author} {\bibfnamefont {A.}~\bibnamefont
  {MacKinnon}},\ }\bibfield  {title} {\bibinfo {title} {Localization: theory
  and experiment},\ }\href {https://doi.org/10.1088/0034-4885/56/12/001}
  {\bibfield  {journal} {\bibinfo  {journal} {Reports on Progress in Physics}\
  }\textbf {\bibinfo {volume} {56}},\ \bibinfo {pages} {1469} (\bibinfo {year}
  {1993})}\BibitemShut {NoStop}%
\bibitem [{\citenamefont {Brouwer}\ \emph {et~al.}(1997)\citenamefont
  {Brouwer}, \citenamefont {Silvestrov},\ and\ \citenamefont
  {Beenakker}}]{brouwer1997}%
  \BibitemOpen
  \bibfield  {author} {\bibinfo {author} {\bibfnamefont {P.~W.}\ \bibnamefont
  {Brouwer}}, \bibinfo {author} {\bibfnamefont {P.~G.}\ \bibnamefont
  {Silvestrov}},\ and\ \bibinfo {author} {\bibfnamefont {C.~W.~J.}\
  \bibnamefont {Beenakker}},\ }\bibfield  {title} {\bibinfo {title} {Theory of
  directed localization in one dimension},\ }\href
  {https://doi.org/10.1103/PhysRevB.56.R4333} {\bibfield  {journal} {\bibinfo
  {journal} {Phys. Rev. B}\ }\textbf {\bibinfo {volume} {56}},\ \bibinfo
  {pages} {R4333} (\bibinfo {year} {1997})}\BibitemShut {NoStop}%
\bibitem [{\citenamefont {Evers}\ and\ \citenamefont
  {Mirlin}(2008)}]{evers2008}%
  \BibitemOpen
  \bibfield  {author} {\bibinfo {author} {\bibfnamefont {F.}~\bibnamefont
  {Evers}}\ and\ \bibinfo {author} {\bibfnamefont {A.~D.}\ \bibnamefont
  {Mirlin}},\ }\bibfield  {title} {\bibinfo {title} {Anderson transitions},\
  }\href {https://doi.org/10.1103/RevModPhys.80.1355} {\bibfield  {journal}
  {\bibinfo  {journal} {Rev. Mod. Phys.}\ }\textbf {\bibinfo {volume} {80}},\
  \bibinfo {pages} {1355} (\bibinfo {year} {2008})}\BibitemShut {NoStop}%
\bibitem [{\citenamefont {Klitzing}\ \emph {et~al.}(1980)\citenamefont
  {Klitzing}, \citenamefont {Dorda},\ and\ \citenamefont {Pepper}}]{quhall}%
  \BibitemOpen
  \bibfield  {author} {\bibinfo {author} {\bibfnamefont {K.~v.}\ \bibnamefont
  {Klitzing}}, \bibinfo {author} {\bibfnamefont {G.}~\bibnamefont {Dorda}},\
  and\ \bibinfo {author} {\bibfnamefont {M.}~\bibnamefont {Pepper}},\
  }\bibfield  {title} {\bibinfo {title} {New method for high-accuracy
  determination of the fine-structure constant based on quantized hall
  resistance},\ }\href {https://doi.org/10.1103/PhysRevLett.45.494} {\bibfield
  {journal} {\bibinfo  {journal} {Phys. Rev. Lett.}\ }\textbf {\bibinfo
  {volume} {45}},\ \bibinfo {pages} {494} (\bibinfo {year} {1980})}\BibitemShut
  {NoStop}%
\bibitem [{\citenamefont {Laughlin}(1981)}]{laughlin1981}%
  \BibitemOpen
  \bibfield  {author} {\bibinfo {author} {\bibfnamefont {R.~B.}\ \bibnamefont
  {Laughlin}},\ }\bibfield  {title} {\bibinfo {title} {Quantized hall
  conductivity in two dimensions},\ }\href
  {https://doi.org/10.1103/PhysRevB.23.5632} {\bibfield  {journal} {\bibinfo
  {journal} {Phys. Rev. B}\ }\textbf {\bibinfo {volume} {23}},\ \bibinfo
  {pages} {5632} (\bibinfo {year} {1981})}\BibitemShut {NoStop}%
\bibitem [{\citenamefont {Thouless}(1983)}]{thouless1983}%
  \BibitemOpen
  \bibfield  {author} {\bibinfo {author} {\bibfnamefont {D.~J.}\ \bibnamefont
  {Thouless}},\ }\bibfield  {title} {\bibinfo {title} {Quantization of particle
  transport},\ }\href {https://doi.org/10.1103/PhysRevB.27.6083} {\bibfield
  {journal} {\bibinfo  {journal} {Phys. Rev. B}\ }\textbf {\bibinfo {volume}
  {27}},\ \bibinfo {pages} {6083} (\bibinfo {year} {1983})}\BibitemShut
  {NoStop}%
\bibitem [{\citenamefont {Niu}\ and\ \citenamefont
  {Thouless}(1984)}]{niu198401}%
  \BibitemOpen
  \bibfield  {author} {\bibinfo {author} {\bibfnamefont {Q.}~\bibnamefont
  {Niu}}\ and\ \bibinfo {author} {\bibfnamefont {D.~J.}\ \bibnamefont
  {Thouless}},\ }\bibfield  {title} {\bibinfo {title} {Quantised adiabatic
  charge transport in the presence of substrate disorder and many-body
  interaction},\ }\href {https://doi.org/10.1088/0305-4470/17/12/016}
  {\bibfield  {journal} {\bibinfo  {journal} {Journal of Physics A:
  Mathematical and General}\ }\textbf {\bibinfo {volume} {17}},\ \bibinfo
  {pages} {2453} (\bibinfo {year} {1984})}\BibitemShut {NoStop}%
\bibitem [{\citenamefont {Niu}\ \emph {et~al.}(1985)\citenamefont {Niu},
  \citenamefont {Thouless},\ and\ \citenamefont {Wu}}]{niu1985}%
  \BibitemOpen
  \bibfield  {author} {\bibinfo {author} {\bibfnamefont {Q.}~\bibnamefont
  {Niu}}, \bibinfo {author} {\bibfnamefont {D.~J.}\ \bibnamefont {Thouless}},\
  and\ \bibinfo {author} {\bibfnamefont {Y.-S.}\ \bibnamefont {Wu}},\
  }\bibfield  {title} {\bibinfo {title} {Quantized hall conductance as a
  topological invariant},\ }\href {https://doi.org/10.1103/PhysRevB.31.3372}
  {\bibfield  {journal} {\bibinfo  {journal} {Phys. Rev. B}\ }\textbf {\bibinfo
  {volume} {31}},\ \bibinfo {pages} {3372} (\bibinfo {year}
  {1985})}\BibitemShut {NoStop}%
\bibitem [{\citenamefont {Hatsugai}(1993{\natexlab{a}})}]{hatsugai199301}%
  \BibitemOpen
  \bibfield  {author} {\bibinfo {author} {\bibfnamefont {Y.}~\bibnamefont
  {Hatsugai}},\ }\bibfield  {title} {\bibinfo {title} {Chern number and edge
  states in the integer quantum hall effect},\ }\href
  {https://doi.org/10.1103/PhysRevLett.71.3697} {\bibfield  {journal} {\bibinfo
   {journal} {Phys. Rev. Lett.}\ }\textbf {\bibinfo {volume} {71}},\ \bibinfo
  {pages} {3697} (\bibinfo {year} {1993}{\natexlab{a}})}\BibitemShut {NoStop}%
\bibitem [{\citenamefont {Hatsugai}(1993{\natexlab{b}})}]{hatsugai199302}%
  \BibitemOpen
  \bibfield  {author} {\bibinfo {author} {\bibfnamefont {Y.}~\bibnamefont
  {Hatsugai}},\ }\bibfield  {title} {\bibinfo {title} {Edge states in the
  integer quantum hall effect and the riemann surface of the bloch function},\
  }\href {https://doi.org/10.1103/PhysRevB.48.11851} {\bibfield  {journal}
  {\bibinfo  {journal} {Phys. Rev. B}\ }\textbf {\bibinfo {volume} {48}},\
  \bibinfo {pages} {11851} (\bibinfo {year} {1993}{\natexlab{b}})}\BibitemShut
  {NoStop}%
\bibitem [{\citenamefont {Anderson}(1958)}]{anderson1958}%
  \BibitemOpen
  \bibfield  {author} {\bibinfo {author} {\bibfnamefont {P.~W.}\ \bibnamefont
  {Anderson}},\ }\bibfield  {title} {\bibinfo {title} {Absence of diffusion in
  certain random lattices},\ }\href {https://doi.org/10.1103/PhysRev.109.1492}
  {\bibfield  {journal} {\bibinfo  {journal} {Phys. Rev.}\ }\textbf {\bibinfo
  {volume} {109}},\ \bibinfo {pages} {1492} (\bibinfo {year}
  {1958})}\BibitemShut {NoStop}%
\bibitem [{\citenamefont {Abrahams}\ \emph {et~al.}(1979)\citenamefont
  {Abrahams}, \citenamefont {Anderson}, \citenamefont {Licciardello},\ and\
  \citenamefont {Ramakrishnan}}]{anderson1979}%
  \BibitemOpen
  \bibfield  {author} {\bibinfo {author} {\bibfnamefont {E.}~\bibnamefont
  {Abrahams}}, \bibinfo {author} {\bibfnamefont {P.~W.}\ \bibnamefont
  {Anderson}}, \bibinfo {author} {\bibfnamefont {D.~C.}\ \bibnamefont
  {Licciardello}},\ and\ \bibinfo {author} {\bibfnamefont {T.~V.}\ \bibnamefont
  {Ramakrishnan}},\ }\bibfield  {title} {\bibinfo {title} {Scaling theory of
  localization: Absence of quantum diffusion in two dimensions},\ }\href
  {https://doi.org/10.1103/PhysRevLett.42.673} {\bibfield  {journal} {\bibinfo
  {journal} {Phys. Rev. Lett.}\ }\textbf {\bibinfo {volume} {42}},\ \bibinfo
  {pages} {673} (\bibinfo {year} {1979})}\BibitemShut {NoStop}%
\bibitem [{\citenamefont {Mott}\ and\ \citenamefont
  {Peierls}(1937)}]{mott_1937}%
  \BibitemOpen
  \bibfield  {author} {\bibinfo {author} {\bibfnamefont {N.~F.}\ \bibnamefont
  {Mott}}\ and\ \bibinfo {author} {\bibfnamefont {R.}~\bibnamefont {Peierls}},\
  }\bibfield  {title} {\bibinfo {title} {Discussion of the paper by de boer and
  verwey},\ }\href {https://doi.org/10.1088/0959-5309/49/4S/308} {\bibfield
  {journal} {\bibinfo  {journal} {Proceedings of the Physical Society}\
  }\textbf {\bibinfo {volume} {49}},\ \bibinfo {pages} {72} (\bibinfo {year}
  {1937})}\BibitemShut {NoStop}%
\bibitem [{\citenamefont {Mott}(1949)}]{mott_1949}%
  \BibitemOpen
  \bibfield  {author} {\bibinfo {author} {\bibfnamefont {N.~F.}\ \bibnamefont
  {Mott}},\ }\bibfield  {title} {\bibinfo {title} {The basis of the electron
  theory of metals, with special reference to the transition metals},\ }\href
  {https://doi.org/10.1088/0370-1298/62/7/303} {\bibfield  {journal} {\bibinfo
  {journal} {Proceedings of the Physical Society. Section A}\ }\textbf
  {\bibinfo {volume} {62}},\ \bibinfo {pages} {416} (\bibinfo {year}
  {1949})}\BibitemShut {NoStop}%
\bibitem [{\citenamefont {Pan}\ and\ \citenamefont
  {Das~Sarma}(2021)}]{pan2021}%
  \BibitemOpen
  \bibfield  {author} {\bibinfo {author} {\bibfnamefont {H.}~\bibnamefont
  {Pan}}\ and\ \bibinfo {author} {\bibfnamefont {S.}~\bibnamefont
  {Das~Sarma}},\ }\bibfield  {title} {\bibinfo {title} {Interaction-driven
  filling-induced metal-insulator transitions in 2d moir\'e lattices},\ }\href
  {https://doi.org/10.1103/PhysRevLett.127.096802} {\bibfield  {journal}
  {\bibinfo  {journal} {Phys. Rev. Lett.}\ }\textbf {\bibinfo {volume} {127}},\
  \bibinfo {pages} {096802} (\bibinfo {year} {2021})}\BibitemShut {NoStop}%
\bibitem [{\citenamefont {Thouless}\ \emph {et~al.}(1982)\citenamefont
  {Thouless}, \citenamefont {Kohmoto}, \citenamefont {Nightingale},\ and\
  \citenamefont {den Nijs}}]{tknn}%
  \BibitemOpen
  \bibfield  {author} {\bibinfo {author} {\bibfnamefont {D.~J.}\ \bibnamefont
  {Thouless}}, \bibinfo {author} {\bibfnamefont {M.}~\bibnamefont {Kohmoto}},
  \bibinfo {author} {\bibfnamefont {M.~P.}\ \bibnamefont {Nightingale}},\ and\
  \bibinfo {author} {\bibfnamefont {M.}~\bibnamefont {den Nijs}},\ }\bibfield
  {title} {\bibinfo {title} {Quantized hall conductance in a two-dimensional
  periodic potential},\ }\href {https://doi.org/10.1103/PhysRevLett.49.405}
  {\bibfield  {journal} {\bibinfo  {journal} {Phys. Rev. Lett.}\ }\textbf
  {\bibinfo {volume} {49}},\ \bibinfo {pages} {405} (\bibinfo {year}
  {1982})}\BibitemShut {NoStop}%
\bibitem [{\citenamefont {Haldane}(1988)}]{haldane1988}%
  \BibitemOpen
  \bibfield  {author} {\bibinfo {author} {\bibfnamefont {F.~D.~M.}\
  \bibnamefont {Haldane}},\ }\bibfield  {title} {\bibinfo {title} {Model for a
  quantum hall effect without landau levels: Condensed-matter realization of
  the "parity anomaly"},\ }\href {https://doi.org/10.1103/PhysRevLett.61.2015}
  {\bibfield  {journal} {\bibinfo  {journal} {Phys. Rev. Lett.}\ }\textbf
  {\bibinfo {volume} {61}},\ \bibinfo {pages} {2015} (\bibinfo {year}
  {1988})}\BibitemShut {NoStop}%
\bibitem [{\citenamefont {Kane}\ and\ \citenamefont
  {Mele}(2005{\natexlab{a}})}]{kane200501}%
  \BibitemOpen
  \bibfield  {author} {\bibinfo {author} {\bibfnamefont {C.~L.}\ \bibnamefont
  {Kane}}\ and\ \bibinfo {author} {\bibfnamefont {E.~J.}\ \bibnamefont
  {Mele}},\ }\bibfield  {title} {\bibinfo {title} {${Z}_{2}$ topological order
  and the quantum spin hall effect},\ }\href
  {https://doi.org/10.1103/PhysRevLett.95.146802} {\bibfield  {journal}
  {\bibinfo  {journal} {Phys. Rev. Lett.}\ }\textbf {\bibinfo {volume} {95}},\
  \bibinfo {pages} {146802} (\bibinfo {year} {2005}{\natexlab{a}})}\BibitemShut
  {NoStop}%
\bibitem [{\citenamefont {Kane}\ and\ \citenamefont
  {Mele}(2005{\natexlab{b}})}]{kane200502}%
  \BibitemOpen
  \bibfield  {author} {\bibinfo {author} {\bibfnamefont {C.~L.}\ \bibnamefont
  {Kane}}\ and\ \bibinfo {author} {\bibfnamefont {E.~J.}\ \bibnamefont
  {Mele}},\ }\bibfield  {title} {\bibinfo {title} {Quantum spin hall effect in
  graphene},\ }\href {https://doi.org/10.1103/PhysRevLett.95.226801} {\bibfield
   {journal} {\bibinfo  {journal} {Phys. Rev. Lett.}\ }\textbf {\bibinfo
  {volume} {95}},\ \bibinfo {pages} {226801} (\bibinfo {year}
  {2005}{\natexlab{b}})}\BibitemShut {NoStop}%
\bibitem [{\citenamefont {Bernevig}\ \emph {et~al.}(2006)\citenamefont
  {Bernevig}, \citenamefont {Hughes},\ and\ \citenamefont
  {Zhang}}]{quspinhallwell}%
  \BibitemOpen
  \bibfield  {author} {\bibinfo {author} {\bibfnamefont {B.~A.}\ \bibnamefont
  {Bernevig}}, \bibinfo {author} {\bibfnamefont {T.~L.}\ \bibnamefont
  {Hughes}},\ and\ \bibinfo {author} {\bibfnamefont {S.-C.}\ \bibnamefont
  {Zhang}},\ }\bibfield  {title} {\bibinfo {title} {Quantum spin hall effect
  and topological phase transition in hgte quantum wells},\ }\href
  {https://doi.org/10.1126/science.1133734} {\bibfield  {journal} {\bibinfo
  {journal} {Science}\ }\textbf {\bibinfo {volume} {314}},\ \bibinfo {pages}
  {1757} (\bibinfo {year} {2006})}\BibitemShut {NoStop}%
\bibitem [{\citenamefont {Fu}\ \emph {et~al.}(2007)\citenamefont {Fu},
  \citenamefont {Kane},\ and\ \citenamefont {Mele}}]{fu2007}%
  \BibitemOpen
  \bibfield  {author} {\bibinfo {author} {\bibfnamefont {L.}~\bibnamefont
  {Fu}}, \bibinfo {author} {\bibfnamefont {C.~L.}\ \bibnamefont {Kane}},\ and\
  \bibinfo {author} {\bibfnamefont {E.~J.}\ \bibnamefont {Mele}},\ }\bibfield
  {title} {\bibinfo {title} {Topological insulators in three dimensions},\
  }\href {https://doi.org/10.1103/PhysRevLett.98.106803} {\bibfield  {journal}
  {\bibinfo  {journal} {Phys. Rev. Lett.}\ }\textbf {\bibinfo {volume} {98}},\
  \bibinfo {pages} {106803} (\bibinfo {year} {2007})}\BibitemShut {NoStop}%
\bibitem [{\citenamefont {Castro~Neto}\ \emph {et~al.}(2009)\citenamefont
  {Castro~Neto}, \citenamefont {Guinea}, \citenamefont {Peres}, \citenamefont
  {Novoselov},\ and\ \citenamefont {Geim}}]{castro2009}%
  \BibitemOpen
  \bibfield  {author} {\bibinfo {author} {\bibfnamefont {A.~H.}\ \bibnamefont
  {Castro~Neto}}, \bibinfo {author} {\bibfnamefont {F.}~\bibnamefont {Guinea}},
  \bibinfo {author} {\bibfnamefont {N.~M.~R.}\ \bibnamefont {Peres}}, \bibinfo
  {author} {\bibfnamefont {K.~S.}\ \bibnamefont {Novoselov}},\ and\ \bibinfo
  {author} {\bibfnamefont {A.~K.}\ \bibnamefont {Geim}},\ }\bibfield  {title}
  {\bibinfo {title} {The electronic properties of graphene},\ }\href
  {https://doi.org/10.1103/RevModPhys.81.109} {\bibfield  {journal} {\bibinfo
  {journal} {Rev. Mod. Phys.}\ }\textbf {\bibinfo {volume} {81}},\ \bibinfo
  {pages} {109} (\bibinfo {year} {2009})}\BibitemShut {NoStop}%
\bibitem [{\citenamefont {Kitaev}(2009)}]{kitaev2009}%
  \BibitemOpen
  \bibfield  {author} {\bibinfo {author} {\bibfnamefont {A.}~\bibnamefont
  {Kitaev}},\ }\bibfield  {title} {\bibinfo {title} {{Periodic table for
  topological insulators and superconductors}},\ }\href
  {https://doi.org/10.1063/1.3149495} {\bibfield  {journal} {\bibinfo
  {journal} {AIP Conference Proceedings}\ }\textbf {\bibinfo {volume} {1134}},\
  \bibinfo {pages} {22} (\bibinfo {year} {2009})}\BibitemShut {NoStop}%
\bibitem [{\citenamefont {Hasan}\ and\ \citenamefont
  {Kane}(2010)}]{hassan2010}%
  \BibitemOpen
  \bibfield  {author} {\bibinfo {author} {\bibfnamefont {M.~Z.}\ \bibnamefont
  {Hasan}}\ and\ \bibinfo {author} {\bibfnamefont {C.~L.}\ \bibnamefont
  {Kane}},\ }\bibfield  {title} {\bibinfo {title} {Colloquium: Topological
  insulators},\ }\href {https://doi.org/10.1103/RevModPhys.82.3045} {\bibfield
  {journal} {\bibinfo  {journal} {Rev. Mod. Phys.}\ }\textbf {\bibinfo {volume}
  {82}},\ \bibinfo {pages} {3045} (\bibinfo {year} {2010})}\BibitemShut
  {NoStop}%
\bibitem [{\citenamefont {Qi}\ and\ \citenamefont {Zhang}(2011)}]{qi2011}%
  \BibitemOpen
  \bibfield  {author} {\bibinfo {author} {\bibfnamefont {X.-L.}\ \bibnamefont
  {Qi}}\ and\ \bibinfo {author} {\bibfnamefont {S.-C.}\ \bibnamefont {Zhang}},\
  }\bibfield  {title} {\bibinfo {title} {Topological insulators and
  superconductors},\ }\href {https://doi.org/10.1103/RevModPhys.83.1057}
  {\bibfield  {journal} {\bibinfo  {journal} {Rev. Mod. Phys.}\ }\textbf
  {\bibinfo {volume} {83}},\ \bibinfo {pages} {1057} (\bibinfo {year}
  {2011})}\BibitemShut {NoStop}%
\bibitem [{\citenamefont {Chiu}\ \emph {et~al.}(2016)\citenamefont {Chiu},
  \citenamefont {Teo}, \citenamefont {Schnyder},\ and\ \citenamefont
  {Ryu}}]{chiu2016}%
  \BibitemOpen
  \bibfield  {author} {\bibinfo {author} {\bibfnamefont {C.-K.}\ \bibnamefont
  {Chiu}}, \bibinfo {author} {\bibfnamefont {J.~C.~Y.}\ \bibnamefont {Teo}},
  \bibinfo {author} {\bibfnamefont {A.~P.}\ \bibnamefont {Schnyder}},\ and\
  \bibinfo {author} {\bibfnamefont {S.}~\bibnamefont {Ryu}},\ }\bibfield
  {title} {\bibinfo {title} {Classification of topological quantum matter with
  symmetries},\ }\href {https://doi.org/10.1103/RevModPhys.88.035005}
  {\bibfield  {journal} {\bibinfo  {journal} {Rev. Mod. Phys.}\ }\textbf
  {\bibinfo {volume} {88}},\ \bibinfo {pages} {035005} (\bibinfo {year}
  {2016})}\BibitemShut {NoStop}%
\bibitem [{\citenamefont {Rotter}(2009)}]{open1}%
  \BibitemOpen
  \bibfield  {author} {\bibinfo {author} {\bibfnamefont {I.}~\bibnamefont
  {Rotter}},\ }\bibfield  {title} {\bibinfo {title} {A non-hermitian hamilton
  operator and the physics of open quantum systems},\ }\href
  {https://doi.org/10.1088/1751-8113/42/15/153001} {\bibfield  {journal}
  {\bibinfo  {journal} {Journal of Physics A: Mathematical and Theoretical}\
  }\textbf {\bibinfo {volume} {42}},\ \bibinfo {pages} {153001} (\bibinfo
  {year} {2009})}\BibitemShut {NoStop}%
\bibitem [{\citenamefont {Malzard}\ \emph {et~al.}(2015)\citenamefont
  {Malzard}, \citenamefont {Poli},\ and\ \citenamefont {Schomerus}}]{open2}%
  \BibitemOpen
  \bibfield  {author} {\bibinfo {author} {\bibfnamefont {S.}~\bibnamefont
  {Malzard}}, \bibinfo {author} {\bibfnamefont {C.}~\bibnamefont {Poli}},\ and\
  \bibinfo {author} {\bibfnamefont {H.}~\bibnamefont {Schomerus}},\ }\bibfield
  {title} {\bibinfo {title} {Topologically protected defect states in open
  photonic systems with non-hermitian charge-conjugation and parity-time
  symmetry},\ }\href {https://doi.org/10.1103/PhysRevLett.115.200402}
  {\bibfield  {journal} {\bibinfo  {journal} {Phys. Rev. Lett.}\ }\textbf
  {\bibinfo {volume} {115}},\ \bibinfo {pages} {200402} (\bibinfo {year}
  {2015})}\BibitemShut {NoStop}%
\bibitem [{\citenamefont {Carmichael}(1993)}]{open3}%
  \BibitemOpen
  \bibfield  {author} {\bibinfo {author} {\bibfnamefont {H.~J.}\ \bibnamefont
  {Carmichael}},\ }\bibfield  {title} {\bibinfo {title} {Quantum trajectory
  theory for cascaded open systems},\ }\href
  {https://doi.org/10.1103/PhysRevLett.70.2273} {\bibfield  {journal} {\bibinfo
   {journal} {Phys. Rev. Lett.}\ }\textbf {\bibinfo {volume} {70}},\ \bibinfo
  {pages} {2273} (\bibinfo {year} {1993})}\BibitemShut {NoStop}%
\bibitem [{\citenamefont {Guo}\ and\ \citenamefont {Poletti}(2017)}]{open4}%
  \BibitemOpen
  \bibfield  {author} {\bibinfo {author} {\bibfnamefont {C.}~\bibnamefont
  {Guo}}\ and\ \bibinfo {author} {\bibfnamefont {D.}~\bibnamefont {Poletti}},\
  }\bibfield  {title} {\bibinfo {title} {Solutions for bosonic and fermionic
  dissipative quadratic open systems},\ }\href
  {https://doi.org/10.1103/PhysRevA.95.052107} {\bibfield  {journal} {\bibinfo
  {journal} {Phys. Rev. A}\ }\textbf {\bibinfo {volume} {95}},\ \bibinfo
  {pages} {052107} (\bibinfo {year} {2017})}\BibitemShut {NoStop}%
\bibitem [{\citenamefont {Dangel}\ \emph {et~al.}(2018)\citenamefont {Dangel},
  \citenamefont {Wagner}, \citenamefont {Cartarius}, \citenamefont {Main},\
  and\ \citenamefont {Wunner}}]{open5}%
  \BibitemOpen
  \bibfield  {author} {\bibinfo {author} {\bibfnamefont {F.}~\bibnamefont
  {Dangel}}, \bibinfo {author} {\bibfnamefont {M.}~\bibnamefont {Wagner}},
  \bibinfo {author} {\bibfnamefont {H.}~\bibnamefont {Cartarius}}, \bibinfo
  {author} {\bibfnamefont {J.}~\bibnamefont {Main}},\ and\ \bibinfo {author}
  {\bibfnamefont {G.}~\bibnamefont {Wunner}},\ }\bibfield  {title} {\bibinfo
  {title} {Topological invariants in dissipative extensions of the
  su-schrieffer-heeger model},\ }\href
  {https://doi.org/10.1103/PhysRevA.98.013628} {\bibfield  {journal} {\bibinfo
  {journal} {Phys. Rev. A}\ }\textbf {\bibinfo {volume} {98}},\ \bibinfo
  {pages} {013628} (\bibinfo {year} {2018})}\BibitemShut {NoStop}%
\bibitem [{\citenamefont {Song}\ \emph {et~al.}(2019)\citenamefont {Song},
  \citenamefont {Yao},\ and\ \citenamefont {Wang}}]{open6}%
  \BibitemOpen
  \bibfield  {author} {\bibinfo {author} {\bibfnamefont {F.}~\bibnamefont
  {Song}}, \bibinfo {author} {\bibfnamefont {S.}~\bibnamefont {Yao}},\ and\
  \bibinfo {author} {\bibfnamefont {Z.}~\bibnamefont {Wang}},\ }\bibfield
  {title} {\bibinfo {title} {Non-hermitian skin effect and chiral damping in
  open quantum systems},\ }\href
  {https://doi.org/10.1103/PhysRevLett.123.170401} {\bibfield  {journal}
  {\bibinfo  {journal} {Phys. Rev. Lett.}\ }\textbf {\bibinfo {volume} {123}},\
  \bibinfo {pages} {170401} (\bibinfo {year} {2019})}\BibitemShut {NoStop}%
\bibitem [{\citenamefont {McDonald}\ \emph {et~al.}(2022)\citenamefont
  {McDonald}, \citenamefont {Hanai},\ and\ \citenamefont {Clerk}}]{open7}%
  \BibitemOpen
  \bibfield  {author} {\bibinfo {author} {\bibfnamefont {A.}~\bibnamefont
  {McDonald}}, \bibinfo {author} {\bibfnamefont {R.}~\bibnamefont {Hanai}},\
  and\ \bibinfo {author} {\bibfnamefont {A.~A.}\ \bibnamefont {Clerk}},\
  }\bibfield  {title} {\bibinfo {title} {Nonequilibrium stationary states of
  quantum non-hermitian lattice models},\ }\href
  {https://doi.org/10.1103/PhysRevB.105.064302} {\bibfield  {journal} {\bibinfo
   {journal} {Phys. Rev. B}\ }\textbf {\bibinfo {volume} {105}},\ \bibinfo
  {pages} {064302} (\bibinfo {year} {2022})}\BibitemShut {NoStop}%
\bibitem [{\citenamefont {Altland}\ \emph {et~al.}(2021)\citenamefont
  {Altland}, \citenamefont {Fleischhauer},\ and\ \citenamefont
  {Diehl}}]{open8}%
  \BibitemOpen
  \bibfield  {author} {\bibinfo {author} {\bibfnamefont {A.}~\bibnamefont
  {Altland}}, \bibinfo {author} {\bibfnamefont {M.}~\bibnamefont
  {Fleischhauer}},\ and\ \bibinfo {author} {\bibfnamefont {S.}~\bibnamefont
  {Diehl}},\ }\bibfield  {title} {\bibinfo {title} {Symmetry classes of open
  fermionic quantum matter},\ }\href
  {https://doi.org/10.1103/PhysRevX.11.021037} {\bibfield  {journal} {\bibinfo
  {journal} {Phys. Rev. X}\ }\textbf {\bibinfo {volume} {11}},\ \bibinfo
  {pages} {021037} (\bibinfo {year} {2021})}\BibitemShut {NoStop}%
\bibitem [{\citenamefont {Guo}\ \emph {et~al.}(2009)\citenamefont {Guo},
  \citenamefont {Salamo}, \citenamefont {Duchesne}, \citenamefont {Morandotti},
  \citenamefont {Volatier-Ravat}, \citenamefont {Aimez}, \citenamefont
  {Siviloglou},\ and\ \citenamefont {Christodoulides}}]{optical1}%
  \BibitemOpen
  \bibfield  {author} {\bibinfo {author} {\bibfnamefont {A.}~\bibnamefont
  {Guo}}, \bibinfo {author} {\bibfnamefont {G.~J.}\ \bibnamefont {Salamo}},
  \bibinfo {author} {\bibfnamefont {D.}~\bibnamefont {Duchesne}}, \bibinfo
  {author} {\bibfnamefont {R.}~\bibnamefont {Morandotti}}, \bibinfo {author}
  {\bibfnamefont {M.}~\bibnamefont {Volatier-Ravat}}, \bibinfo {author}
  {\bibfnamefont {V.}~\bibnamefont {Aimez}}, \bibinfo {author} {\bibfnamefont
  {G.~A.}\ \bibnamefont {Siviloglou}},\ and\ \bibinfo {author} {\bibfnamefont
  {D.~N.}\ \bibnamefont {Christodoulides}},\ }\bibfield  {title} {\bibinfo
  {title} {Observation of $\mathcal{P}\mathcal{T}$-symmetry breaking in complex
  optical potentials},\ }\href {https://doi.org/10.1103/PhysRevLett.103.093902}
  {\bibfield  {journal} {\bibinfo  {journal} {Phys. Rev. Lett.}\ }\textbf
  {\bibinfo {volume} {103}},\ \bibinfo {pages} {093902} (\bibinfo {year}
  {2009})}\BibitemShut {NoStop}%
\bibitem [{\citenamefont {Chen}\ \emph {et~al.}(2017)\citenamefont {Chen},
  \citenamefont {Kaya~{\"O}zdemir}, \citenamefont {Zhao}, \citenamefont
  {Wiersig},\ and\ \citenamefont {Yang}}]{optical2}%
  \BibitemOpen
  \bibfield  {author} {\bibinfo {author} {\bibfnamefont {W.}~\bibnamefont
  {Chen}}, \bibinfo {author} {\bibfnamefont {{\c{S}}.}~\bibnamefont
  {Kaya~{\"O}zdemir}}, \bibinfo {author} {\bibfnamefont {G.}~\bibnamefont
  {Zhao}}, \bibinfo {author} {\bibfnamefont {J.}~\bibnamefont {Wiersig}},\ and\
  \bibinfo {author} {\bibfnamefont {L.}~\bibnamefont {Yang}},\ }\bibfield
  {title} {\bibinfo {title} {Exceptional points enhance sensing in an optical
  microcavity},\ }\href {https://doi.org/10.1038/nature23281} {\bibfield
  {journal} {\bibinfo  {journal} {Nature}\ }\textbf {\bibinfo {volume} {548}},\
  \bibinfo {pages} {192} (\bibinfo {year} {2017})}\BibitemShut {NoStop}%
\bibitem [{\citenamefont {Xiao}\ \emph {et~al.}(2019)\citenamefont {Xiao},
  \citenamefont {Wang}, \citenamefont {Zhan}, \citenamefont {Bian},
  \citenamefont {Kawabata}, \citenamefont {Ueda}, \citenamefont {Yi},\ and\
  \citenamefont {Xue}}]{optical3}%
  \BibitemOpen
  \bibfield  {author} {\bibinfo {author} {\bibfnamefont {L.}~\bibnamefont
  {Xiao}}, \bibinfo {author} {\bibfnamefont {K.}~\bibnamefont {Wang}}, \bibinfo
  {author} {\bibfnamefont {X.}~\bibnamefont {Zhan}}, \bibinfo {author}
  {\bibfnamefont {Z.}~\bibnamefont {Bian}}, \bibinfo {author} {\bibfnamefont
  {K.}~\bibnamefont {Kawabata}}, \bibinfo {author} {\bibfnamefont
  {M.}~\bibnamefont {Ueda}}, \bibinfo {author} {\bibfnamefont {W.}~\bibnamefont
  {Yi}},\ and\ \bibinfo {author} {\bibfnamefont {P.}~\bibnamefont {Xue}},\
  }\bibfield  {title} {\bibinfo {title} {Observation of critical phenomena in
  parity-time-symmetric quantum dynamics},\ }\href
  {https://doi.org/10.1103/PhysRevLett.123.230401} {\bibfield  {journal}
  {\bibinfo  {journal} {Phys. Rev. Lett.}\ }\textbf {\bibinfo {volume} {123}},\
  \bibinfo {pages} {230401} (\bibinfo {year} {2019})}\BibitemShut {NoStop}%
\bibitem [{\citenamefont {Xiao}\ \emph {et~al.}(2020)\citenamefont {Xiao},
  \citenamefont {Deng}, \citenamefont {Wang}, \citenamefont {Zhu},
  \citenamefont {Wang}, \citenamefont {Yi},\ and\ \citenamefont
  {Xue}}]{optical4}%
  \BibitemOpen
  \bibfield  {author} {\bibinfo {author} {\bibfnamefont {L.}~\bibnamefont
  {Xiao}}, \bibinfo {author} {\bibfnamefont {T.}~\bibnamefont {Deng}}, \bibinfo
  {author} {\bibfnamefont {K.}~\bibnamefont {Wang}}, \bibinfo {author}
  {\bibfnamefont {G.}~\bibnamefont {Zhu}}, \bibinfo {author} {\bibfnamefont
  {Z.}~\bibnamefont {Wang}}, \bibinfo {author} {\bibfnamefont {W.}~\bibnamefont
  {Yi}},\ and\ \bibinfo {author} {\bibfnamefont {P.}~\bibnamefont {Xue}},\
  }\bibfield  {title} {\bibinfo {title} {Non-hermitian bulk--boundary
  correspondence in quantum dynamics},\ }\href
  {https://doi.org/10.1038/s41567-020-0836-6} {\bibfield  {journal} {\bibinfo
  {journal} {Nature Physics}\ }\textbf {\bibinfo {volume} {16}},\ \bibinfo
  {pages} {761} (\bibinfo {year} {2020})}\BibitemShut {NoStop}%
\bibitem [{\citenamefont {Xiao}\ \emph {et~al.}(2021)\citenamefont {Xiao},
  \citenamefont {Deng}, \citenamefont {Wang}, \citenamefont {Wang},
  \citenamefont {Yi},\ and\ \citenamefont {Xue}}]{optical5}%
  \BibitemOpen
  \bibfield  {author} {\bibinfo {author} {\bibfnamefont {L.}~\bibnamefont
  {Xiao}}, \bibinfo {author} {\bibfnamefont {T.}~\bibnamefont {Deng}}, \bibinfo
  {author} {\bibfnamefont {K.}~\bibnamefont {Wang}}, \bibinfo {author}
  {\bibfnamefont {Z.}~\bibnamefont {Wang}}, \bibinfo {author} {\bibfnamefont
  {W.}~\bibnamefont {Yi}},\ and\ \bibinfo {author} {\bibfnamefont
  {P.}~\bibnamefont {Xue}},\ }\bibfield  {title} {\bibinfo {title} {Observation
  of non-bloch parity-time symmetry and exceptional points},\ }\href
  {https://doi.org/10.1103/PhysRevLett.126.230402} {\bibfield  {journal}
  {\bibinfo  {journal} {Phys. Rev. Lett.}\ }\textbf {\bibinfo {volume} {126}},\
  \bibinfo {pages} {230402} (\bibinfo {year} {2021})}\BibitemShut {NoStop}%
\bibitem [{\citenamefont {Xiao}\ \emph {et~al.}(2023)\citenamefont {Xiao},
  \citenamefont {Xue}, \citenamefont {Song}, \citenamefont {Hu}, \citenamefont
  {Yi}, \citenamefont {Wang},\ and\ \citenamefont {Xue}}]{optical6}%
  \BibitemOpen
  \bibfield  {author} {\bibinfo {author} {\bibfnamefont {L.}~\bibnamefont
  {Xiao}}, \bibinfo {author} {\bibfnamefont {W.-T.}\ \bibnamefont {Xue}},
  \bibinfo {author} {\bibfnamefont {F.}~\bibnamefont {Song}}, \bibinfo {author}
  {\bibfnamefont {Y.-M.}\ \bibnamefont {Hu}}, \bibinfo {author} {\bibfnamefont
  {W.}~\bibnamefont {Yi}}, \bibinfo {author} {\bibfnamefont {Z.}~\bibnamefont
  {Wang}},\ and\ \bibinfo {author} {\bibfnamefont {P.}~\bibnamefont {Xue}},\
  }\href@noop {} {\bibinfo {title} {Observation of non-hermitian edge burst in
  quantum dynamics}} (\bibinfo {year} {2023}),\ \Eprint
  {https://arxiv.org/abs/2303.12831} {arXiv:2303.12831 [cond-mat.mes-hall]}
  \BibitemShut {NoStop}%
\bibitem [{\citenamefont {Ezawa}(2019{\natexlab{a}})}]{circuit1}%
  \BibitemOpen
  \bibfield  {author} {\bibinfo {author} {\bibfnamefont {M.}~\bibnamefont
  {Ezawa}},\ }\bibfield  {title} {\bibinfo {title} {Non-hermitian boundary and
  interface states in nonreciprocal higher-order topological metals and
  electrical circuits},\ }\href {https://doi.org/10.1103/PhysRevB.99.121411}
  {\bibfield  {journal} {\bibinfo  {journal} {Phys. Rev. B}\ }\textbf {\bibinfo
  {volume} {99}},\ \bibinfo {pages} {121411} (\bibinfo {year}
  {2019}{\natexlab{a}})}\BibitemShut {NoStop}%
\bibitem [{\citenamefont {Ezawa}(2019{\natexlab{b}})}]{circuit2}%
  \BibitemOpen
  \bibfield  {author} {\bibinfo {author} {\bibfnamefont {M.}~\bibnamefont
  {Ezawa}},\ }\bibfield  {title} {\bibinfo {title} {Electric circuits for
  non-hermitian chern insulators},\ }\href
  {https://doi.org/10.1103/PhysRevB.100.081401} {\bibfield  {journal} {\bibinfo
   {journal} {Phys. Rev. B}\ }\textbf {\bibinfo {volume} {100}},\ \bibinfo
  {pages} {081401} (\bibinfo {year} {2019}{\natexlab{b}})}\BibitemShut
  {NoStop}%
\bibitem [{\citenamefont {Hofmann}\ \emph {et~al.}(2020)\citenamefont
  {Hofmann}, \citenamefont {Helbig}, \citenamefont {Schindler}, \citenamefont
  {Salgo}, \citenamefont {Brzezi\ifmmode~\acute{n}\else \'{n}\fi{}ska},
  \citenamefont {Greiter}, \citenamefont {Kiessling}, \citenamefont {Wolf},
  \citenamefont {Vollhardt}, \citenamefont {Kaba\ifmmode~\check{s}\else
  \v{s}\fi{}i}, \citenamefont {Lee}, \citenamefont {Bilu\ifmmode \check{s}\else
  \v{s}\fi{}i\ifmmode~\acute{c}\else \'{c}\fi{}}, \citenamefont {Thomale},\
  and\ \citenamefont {Neupert}}]{circuit3}%
  \BibitemOpen
  \bibfield  {author} {\bibinfo {author} {\bibfnamefont {T.}~\bibnamefont
  {Hofmann}}, \bibinfo {author} {\bibfnamefont {T.}~\bibnamefont {Helbig}},
  \bibinfo {author} {\bibfnamefont {F.}~\bibnamefont {Schindler}}, \bibinfo
  {author} {\bibfnamefont {N.}~\bibnamefont {Salgo}}, \bibinfo {author}
  {\bibfnamefont {M.}~\bibnamefont {Brzezi\ifmmode~\acute{n}\else
  \'{n}\fi{}ska}}, \bibinfo {author} {\bibfnamefont {M.}~\bibnamefont
  {Greiter}}, \bibinfo {author} {\bibfnamefont {T.}~\bibnamefont {Kiessling}},
  \bibinfo {author} {\bibfnamefont {D.}~\bibnamefont {Wolf}}, \bibinfo {author}
  {\bibfnamefont {A.}~\bibnamefont {Vollhardt}}, \bibinfo {author}
  {\bibfnamefont {A.}~\bibnamefont {Kaba\ifmmode~\check{s}\else \v{s}\fi{}i}},
  \bibinfo {author} {\bibfnamefont {C.~H.}\ \bibnamefont {Lee}}, \bibinfo
  {author} {\bibfnamefont {A.}~\bibnamefont {Bilu\ifmmode \check{s}\else
  \v{s}\fi{}i\ifmmode~\acute{c}\else \'{c}\fi{}}}, \bibinfo {author}
  {\bibfnamefont {R.}~\bibnamefont {Thomale}},\ and\ \bibinfo {author}
  {\bibfnamefont {T.}~\bibnamefont {Neupert}},\ }\bibfield  {title} {\bibinfo
  {title} {Reciprocal skin effect and its realization in a topolectrical
  circuit},\ }\href {https://doi.org/10.1103/PhysRevResearch.2.023265}
  {\bibfield  {journal} {\bibinfo  {journal} {Phys. Rev. Res.}\ }\textbf
  {\bibinfo {volume} {2}},\ \bibinfo {pages} {023265} (\bibinfo {year}
  {2020})}\BibitemShut {NoStop}%
\bibitem [{\citenamefont {Helbig}\ \emph {et~al.}(2020)\citenamefont {Helbig},
  \citenamefont {Hofmann}, \citenamefont {Imhof}, \citenamefont {Abdelghany},
  \citenamefont {Kiessling}, \citenamefont {Molenkamp}, \citenamefont {Lee},
  \citenamefont {Szameit}, \citenamefont {Greiter},\ and\ \citenamefont
  {Thomale}}]{circuit4}%
  \BibitemOpen
  \bibfield  {author} {\bibinfo {author} {\bibfnamefont {T.}~\bibnamefont
  {Helbig}}, \bibinfo {author} {\bibfnamefont {T.}~\bibnamefont {Hofmann}},
  \bibinfo {author} {\bibfnamefont {S.}~\bibnamefont {Imhof}}, \bibinfo
  {author} {\bibfnamefont {M.}~\bibnamefont {Abdelghany}}, \bibinfo {author}
  {\bibfnamefont {T.}~\bibnamefont {Kiessling}}, \bibinfo {author}
  {\bibfnamefont {L.~W.}\ \bibnamefont {Molenkamp}}, \bibinfo {author}
  {\bibfnamefont {C.~H.}\ \bibnamefont {Lee}}, \bibinfo {author} {\bibfnamefont
  {A.}~\bibnamefont {Szameit}}, \bibinfo {author} {\bibfnamefont
  {M.}~\bibnamefont {Greiter}},\ and\ \bibinfo {author} {\bibfnamefont
  {R.}~\bibnamefont {Thomale}},\ }\bibfield  {title} {\bibinfo {title}
  {Generalized bulk--boundary correspondence in non-hermitian topolectrical
  circuits},\ }\href {https://doi.org/10.1038/s41567-020-0922-9} {\bibfield
  {journal} {\bibinfo  {journal} {Nature Physics}\ }\textbf {\bibinfo {volume}
  {16}},\ \bibinfo {pages} {747} (\bibinfo {year} {2020})}\BibitemShut
  {NoStop}%
\bibitem [{\citenamefont {Li}\ \emph {et~al.}(2021)\citenamefont {Li},
  \citenamefont {Lee},\ and\ \citenamefont {Gong}}]{circuit5}%
  \BibitemOpen
  \bibfield  {author} {\bibinfo {author} {\bibfnamefont {L.}~\bibnamefont
  {Li}}, \bibinfo {author} {\bibfnamefont {C.~H.}\ \bibnamefont {Lee}},\ and\
  \bibinfo {author} {\bibfnamefont {J.}~\bibnamefont {Gong}},\ }\bibfield
  {title} {\bibinfo {title} {Impurity induced scale-free localization},\ }\href
  {https://doi.org/10.1038/s42005-021-00547-x} {\bibfield  {journal} {\bibinfo
  {journal} {Communications Physics}\ }\textbf {\bibinfo {volume} {4}},\
  \bibinfo {pages} {42} (\bibinfo {year} {2021})}\BibitemShut {NoStop}%
\bibitem [{\citenamefont {Hu}\ and\ \citenamefont {Zhao}(2021)}]{circuit6}%
  \BibitemOpen
  \bibfield  {author} {\bibinfo {author} {\bibfnamefont {H.}~\bibnamefont
  {Hu}}\ and\ \bibinfo {author} {\bibfnamefont {E.}~\bibnamefont {Zhao}},\
  }\bibfield  {title} {\bibinfo {title} {Knots and non-hermitian bloch bands},\
  }\href {https://doi.org/10.1103/PhysRevLett.126.010401} {\bibfield  {journal}
  {\bibinfo  {journal} {Phys. Rev. Lett.}\ }\textbf {\bibinfo {volume} {126}},\
  \bibinfo {pages} {010401} (\bibinfo {year} {2021})}\BibitemShut {NoStop}%
\bibitem [{\citenamefont {Bergholtz}\ \emph {et~al.}(2021)\citenamefont
  {Bergholtz}, \citenamefont {Budich},\ and\ \citenamefont
  {Kunst}}]{bergholtzrev2021}%
  \BibitemOpen
  \bibfield  {author} {\bibinfo {author} {\bibfnamefont {E.~J.}\ \bibnamefont
  {Bergholtz}}, \bibinfo {author} {\bibfnamefont {J.~C.}\ \bibnamefont
  {Budich}},\ and\ \bibinfo {author} {\bibfnamefont {F.~K.}\ \bibnamefont
  {Kunst}},\ }\bibfield  {title} {\bibinfo {title} {Exceptional topology of
  non-hermitian systems},\ }\href
  {https://doi.org/10.1103/RevModPhys.93.015005} {\bibfield  {journal}
  {\bibinfo  {journal} {Rev. Mod. Phys.}\ }\textbf {\bibinfo {volume} {93}},\
  \bibinfo {pages} {015005} (\bibinfo {year} {2021})}\BibitemShut {NoStop}%
\bibitem [{\citenamefont {Yuto}\ \emph {et~al.}(2020)\citenamefont {Yuto},
  \citenamefont {Zongping},\ and\ \citenamefont {Masahito}}]{ashida2020}%
  \BibitemOpen
  \bibfield  {author} {\bibinfo {author} {\bibfnamefont {A.}~\bibnamefont
  {Yuto}}, \bibinfo {author} {\bibfnamefont {G.}~\bibnamefont {Zongping}},\
  and\ \bibinfo {author} {\bibfnamefont {U.}~\bibnamefont {Masahito}},\
  }\bibfield  {title} {\bibinfo {title} {Non-hermitian physics},\ }\href
  {https://doi.org/10.1080/00018732.2021.1876991} {\bibfield  {journal}
  {\bibinfo  {journal} {Advances in Physics}\ }\textbf {\bibinfo {volume}
  {69}},\ \bibinfo {pages} {249} (\bibinfo {year} {2020})}\BibitemShut
  {NoStop}%
\bibitem [{\citenamefont {Gong}\ \emph {et~al.}(2018)\citenamefont {Gong},
  \citenamefont {Ashida}, \citenamefont {Kawabata}, \citenamefont {Takasan},
  \citenamefont {Higashikawa},\ and\ \citenamefont {Ueda}}]{gong2018}%
  \BibitemOpen
  \bibfield  {author} {\bibinfo {author} {\bibfnamefont {Z.}~\bibnamefont
  {Gong}}, \bibinfo {author} {\bibfnamefont {Y.}~\bibnamefont {Ashida}},
  \bibinfo {author} {\bibfnamefont {K.}~\bibnamefont {Kawabata}}, \bibinfo
  {author} {\bibfnamefont {K.}~\bibnamefont {Takasan}}, \bibinfo {author}
  {\bibfnamefont {S.}~\bibnamefont {Higashikawa}},\ and\ \bibinfo {author}
  {\bibfnamefont {M.}~\bibnamefont {Ueda}},\ }\bibfield  {title} {\bibinfo
  {title} {Topological phases of non-hermitian systems},\ }\href
  {https://doi.org/10.1103/PhysRevX.8.031079} {\bibfield  {journal} {\bibinfo
  {journal} {Phys. Rev. X}\ }\textbf {\bibinfo {volume} {8}},\ \bibinfo {pages}
  {031079} (\bibinfo {year} {2018})}\BibitemShut {NoStop}%
\bibitem [{\citenamefont {Kawabata}\ \emph {et~al.}(2019)\citenamefont
  {Kawabata}, \citenamefont {Shiozaki}, \citenamefont {Ueda},\ and\
  \citenamefont {Sato}}]{kawabataprx}%
  \BibitemOpen
  \bibfield  {author} {\bibinfo {author} {\bibfnamefont {K.}~\bibnamefont
  {Kawabata}}, \bibinfo {author} {\bibfnamefont {K.}~\bibnamefont {Shiozaki}},
  \bibinfo {author} {\bibfnamefont {M.}~\bibnamefont {Ueda}},\ and\ \bibinfo
  {author} {\bibfnamefont {M.}~\bibnamefont {Sato}},\ }\bibfield  {title}
  {\bibinfo {title} {Symmetry and topology in non-hermitian physics},\ }\href
  {https://doi.org/10.1103/PhysRevX.9.041015} {\bibfield  {journal} {\bibinfo
  {journal} {Phys. Rev. X}\ }\textbf {\bibinfo {volume} {9}},\ \bibinfo {pages}
  {041015} (\bibinfo {year} {2019})}\BibitemShut {NoStop}%
\bibitem [{\citenamefont {Yao}\ and\ \citenamefont {Wang}(2018)}]{yao2018}%
  \BibitemOpen
  \bibfield  {author} {\bibinfo {author} {\bibfnamefont {S.}~\bibnamefont
  {Yao}}\ and\ \bibinfo {author} {\bibfnamefont {Z.}~\bibnamefont {Wang}},\
  }\bibfield  {title} {\bibinfo {title} {Edge states and topological invariants
  of non-hermitian systems},\ }\href
  {https://doi.org/10.1103/PhysRevLett.121.086803} {\bibfield  {journal}
  {\bibinfo  {journal} {Phys. Rev. Lett.}\ }\textbf {\bibinfo {volume} {121}},\
  \bibinfo {pages} {086803} (\bibinfo {year} {2018})}\BibitemShut {NoStop}%
\bibitem [{\citenamefont {Yokomizo}\ and\ \citenamefont
  {Murakami}(2019)}]{yokomizo2019}%
  \BibitemOpen
  \bibfield  {author} {\bibinfo {author} {\bibfnamefont {K.}~\bibnamefont
  {Yokomizo}}\ and\ \bibinfo {author} {\bibfnamefont {S.}~\bibnamefont
  {Murakami}},\ }\bibfield  {title} {\bibinfo {title} {Non-bloch band theory of
  non-hermitian systems},\ }\href
  {https://doi.org/10.1103/PhysRevLett.123.066404} {\bibfield  {journal}
  {\bibinfo  {journal} {Phys. Rev. Lett.}\ }\textbf {\bibinfo {volume} {123}},\
  \bibinfo {pages} {066404} (\bibinfo {year} {2019})}\BibitemShut {NoStop}%
\bibitem [{\citenamefont {Yang}\ \emph {et~al.}(2020)\citenamefont {Yang},
  \citenamefont {Zhang}, \citenamefont {Fang},\ and\ \citenamefont
  {Hu}}]{yang2020}%
  \BibitemOpen
  \bibfield  {author} {\bibinfo {author} {\bibfnamefont {Z.}~\bibnamefont
  {Yang}}, \bibinfo {author} {\bibfnamefont {K.}~\bibnamefont {Zhang}},
  \bibinfo {author} {\bibfnamefont {C.}~\bibnamefont {Fang}},\ and\ \bibinfo
  {author} {\bibfnamefont {J.}~\bibnamefont {Hu}},\ }\bibfield  {title}
  {\bibinfo {title} {Non-hermitian bulk-boundary correspondence and auxiliary
  generalized brillouin zone theory},\ }\href
  {https://doi.org/10.1103/PhysRevLett.125.226402} {\bibfield  {journal}
  {\bibinfo  {journal} {Phys. Rev. Lett.}\ }\textbf {\bibinfo {volume} {125}},\
  \bibinfo {pages} {226402} (\bibinfo {year} {2020})}\BibitemShut {NoStop}%
\bibitem [{\citenamefont {Zhang}\ \emph {et~al.}(2020)\citenamefont {Zhang},
  \citenamefont {Yang},\ and\ \citenamefont {Fang}}]{zhang2020}%
  \BibitemOpen
  \bibfield  {author} {\bibinfo {author} {\bibfnamefont {K.}~\bibnamefont
  {Zhang}}, \bibinfo {author} {\bibfnamefont {Z.}~\bibnamefont {Yang}},\ and\
  \bibinfo {author} {\bibfnamefont {C.}~\bibnamefont {Fang}},\ }\bibfield
  {title} {\bibinfo {title} {Correspondence between winding numbers and skin
  modes in non-hermitian systems},\ }\href
  {https://doi.org/10.1103/PhysRevLett.125.126402} {\bibfield  {journal}
  {\bibinfo  {journal} {Phys. Rev. Lett.}\ }\textbf {\bibinfo {volume} {125}},\
  \bibinfo {pages} {126402} (\bibinfo {year} {2020})}\BibitemShut {NoStop}%
\bibitem [{\citenamefont {Okuma}\ \emph {et~al.}(2020)\citenamefont {Okuma},
  \citenamefont {Kawabata}, \citenamefont {Shiozaki},\ and\ \citenamefont
  {Sato}}]{origin2020}%
  \BibitemOpen
  \bibfield  {author} {\bibinfo {author} {\bibfnamefont {N.}~\bibnamefont
  {Okuma}}, \bibinfo {author} {\bibfnamefont {K.}~\bibnamefont {Kawabata}},
  \bibinfo {author} {\bibfnamefont {K.}~\bibnamefont {Shiozaki}},\ and\
  \bibinfo {author} {\bibfnamefont {M.}~\bibnamefont {Sato}},\ }\bibfield
  {title} {\bibinfo {title} {Topological origin of non-hermitian skin
  effects},\ }\href {https://doi.org/10.1103/PhysRevLett.124.086801} {\bibfield
   {journal} {\bibinfo  {journal} {Phys. Rev. Lett.}\ }\textbf {\bibinfo
  {volume} {124}},\ \bibinfo {pages} {086801} (\bibinfo {year}
  {2020})}\BibitemShut {NoStop}%
\bibitem [{\citenamefont {Liu}\ \emph {et~al.}(2019)\citenamefont {Liu},
  \citenamefont {Zhang}, \citenamefont {Ai}, \citenamefont {Gong},
  \citenamefont {Kawabata}, \citenamefont {Ueda},\ and\ \citenamefont
  {Nori}}]{liu2019second}%
  \BibitemOpen
  \bibfield  {author} {\bibinfo {author} {\bibfnamefont {T.}~\bibnamefont
  {Liu}}, \bibinfo {author} {\bibfnamefont {Y.-R.}\ \bibnamefont {Zhang}},
  \bibinfo {author} {\bibfnamefont {Q.}~\bibnamefont {Ai}}, \bibinfo {author}
  {\bibfnamefont {Z.}~\bibnamefont {Gong}}, \bibinfo {author} {\bibfnamefont
  {K.}~\bibnamefont {Kawabata}}, \bibinfo {author} {\bibfnamefont
  {M.}~\bibnamefont {Ueda}},\ and\ \bibinfo {author} {\bibfnamefont
  {F.}~\bibnamefont {Nori}},\ }\bibfield  {title} {\bibinfo {title}
  {Second-order topological phases in non-hermitian systems},\ }\href
  {https://doi.org/10.1103/PhysRevLett.122.076801} {\bibfield  {journal}
  {\bibinfo  {journal} {Phys. Rev. Lett.}\ }\textbf {\bibinfo {volume} {122}},\
  \bibinfo {pages} {076801} (\bibinfo {year} {2019})}\BibitemShut {NoStop}%
\bibitem [{\citenamefont {Lee}\ \emph {et~al.}(2019)\citenamefont {Lee},
  \citenamefont {Li},\ and\ \citenamefont {Gong}}]{lee2019ho}%
  \BibitemOpen
  \bibfield  {author} {\bibinfo {author} {\bibfnamefont {C.~H.}\ \bibnamefont
  {Lee}}, \bibinfo {author} {\bibfnamefont {L.}~\bibnamefont {Li}},\ and\
  \bibinfo {author} {\bibfnamefont {J.}~\bibnamefont {Gong}},\ }\bibfield
  {title} {\bibinfo {title} {Hybrid higher-order skin-topological modes in
  nonreciprocal systems},\ }\href
  {https://doi.org/10.1103/PhysRevLett.123.016805} {\bibfield  {journal}
  {\bibinfo  {journal} {Phys. Rev. Lett.}\ }\textbf {\bibinfo {volume} {123}},\
  \bibinfo {pages} {016805} (\bibinfo {year} {2019})}\BibitemShut {NoStop}%
\bibitem [{\citenamefont {Okugawa}\ \emph {et~al.}(2020)\citenamefont
  {Okugawa}, \citenamefont {Takahashi},\ and\ \citenamefont
  {Yokomizo}}]{okugawa2020}%
  \BibitemOpen
  \bibfield  {author} {\bibinfo {author} {\bibfnamefont {R.}~\bibnamefont
  {Okugawa}}, \bibinfo {author} {\bibfnamefont {R.}~\bibnamefont {Takahashi}},\
  and\ \bibinfo {author} {\bibfnamefont {K.}~\bibnamefont {Yokomizo}},\
  }\bibfield  {title} {\bibinfo {title} {Second-order topological non-hermitian
  skin effects},\ }\href {https://doi.org/10.1103/PhysRevB.102.241202}
  {\bibfield  {journal} {\bibinfo  {journal} {Phys. Rev. B}\ }\textbf {\bibinfo
  {volume} {102}},\ \bibinfo {pages} {241202} (\bibinfo {year}
  {2020})}\BibitemShut {NoStop}%
\bibitem [{\citenamefont {Kawabata}\ \emph {et~al.}(2020)\citenamefont
  {Kawabata}, \citenamefont {Sato},\ and\ \citenamefont
  {Shiozaki}}]{kawabatahigher}%
  \BibitemOpen
  \bibfield  {author} {\bibinfo {author} {\bibfnamefont {K.}~\bibnamefont
  {Kawabata}}, \bibinfo {author} {\bibfnamefont {M.}~\bibnamefont {Sato}},\
  and\ \bibinfo {author} {\bibfnamefont {K.}~\bibnamefont {Shiozaki}},\
  }\bibfield  {title} {\bibinfo {title} {Higher-order non-hermitian skin
  effect},\ }\href {https://doi.org/10.1103/PhysRevB.102.205118} {\bibfield
  {journal} {\bibinfo  {journal} {Phys. Rev. B}\ }\textbf {\bibinfo {volume}
  {102}},\ \bibinfo {pages} {205118} (\bibinfo {year} {2020})}\BibitemShut
  {NoStop}%
\bibitem [{\citenamefont {Fu}\ \emph {et~al.}(2021)\citenamefont {Fu},
  \citenamefont {Hu},\ and\ \citenamefont {Wan}}]{fu2021}%
  \BibitemOpen
  \bibfield  {author} {\bibinfo {author} {\bibfnamefont {Y.}~\bibnamefont
  {Fu}}, \bibinfo {author} {\bibfnamefont {J.}~\bibnamefont {Hu}},\ and\
  \bibinfo {author} {\bibfnamefont {S.}~\bibnamefont {Wan}},\ }\bibfield
  {title} {\bibinfo {title} {Non-hermitian second-order skin and topological
  modes},\ }\href {https://doi.org/10.1103/PhysRevB.103.045420} {\bibfield
  {journal} {\bibinfo  {journal} {Phys. Rev. B}\ }\textbf {\bibinfo {volume}
  {103}},\ \bibinfo {pages} {045420} (\bibinfo {year} {2021})}\BibitemShut
  {NoStop}%
\bibitem [{\citenamefont {Li}\ \emph {et~al.}(2022)\citenamefont {Li},
  \citenamefont {Liang}, \citenamefont {Wang}, \citenamefont {Lu},\ and\
  \citenamefont {Liu}}]{st2022}%
  \BibitemOpen
  \bibfield  {author} {\bibinfo {author} {\bibfnamefont {Y.}~\bibnamefont
  {Li}}, \bibinfo {author} {\bibfnamefont {C.}~\bibnamefont {Liang}}, \bibinfo
  {author} {\bibfnamefont {C.}~\bibnamefont {Wang}}, \bibinfo {author}
  {\bibfnamefont {C.}~\bibnamefont {Lu}},\ and\ \bibinfo {author}
  {\bibfnamefont {Y.-C.}\ \bibnamefont {Liu}},\ }\bibfield  {title} {\bibinfo
  {title} {Gain-loss-induced hybrid skin-topological effect},\ }\href
  {https://doi.org/10.1103/PhysRevLett.128.223903} {\bibfield  {journal}
  {\bibinfo  {journal} {Phys. Rev. Lett.}\ }\textbf {\bibinfo {volume} {128}},\
  \bibinfo {pages} {223903} (\bibinfo {year} {2022})}\BibitemShut {NoStop}%
\bibitem [{\citenamefont {Yokomizo}\ and\ \citenamefont
  {Murakami}(2023)}]{yokomizo2023nonbloch}%
  \BibitemOpen
  \bibfield  {author} {\bibinfo {author} {\bibfnamefont {K.}~\bibnamefont
  {Yokomizo}}\ and\ \bibinfo {author} {\bibfnamefont {S.}~\bibnamefont
  {Murakami}},\ }\bibfield  {title} {\bibinfo {title} {Non-bloch bands in
  two-dimensional non-hermitian systems},\ }\href
  {https://doi.org/10.1103/PhysRevB.107.195112} {\bibfield  {journal} {\bibinfo
   {journal} {Phys. Rev. B}\ }\textbf {\bibinfo {volume} {107}},\ \bibinfo
  {pages} {195112} (\bibinfo {year} {2023})}\BibitemShut {NoStop}%
\bibitem [{\citenamefont {Wang}\ \emph
  {et~al.}(2024{\natexlab{a}})\citenamefont {Wang}, \citenamefont {Song},\ and\
  \citenamefont {Wang}}]{wang2024amoeba}%
  \BibitemOpen
  \bibfield  {author} {\bibinfo {author} {\bibfnamefont {H.-Y.}\ \bibnamefont
  {Wang}}, \bibinfo {author} {\bibfnamefont {F.}~\bibnamefont {Song}},\ and\
  \bibinfo {author} {\bibfnamefont {Z.}~\bibnamefont {Wang}},\ }\bibfield
  {title} {\bibinfo {title} {Amoeba formulation of non-bloch band theory in
  arbitrary dimensions},\ }\href {https://doi.org/10.1103/PhysRevX.14.021011}
  {\bibfield  {journal} {\bibinfo  {journal} {Phys. Rev. X}\ }\textbf {\bibinfo
  {volume} {14}},\ \bibinfo {pages} {021011} (\bibinfo {year}
  {2024}{\natexlab{a}})}\BibitemShut {NoStop}%
\bibitem [{\citenamefont {Hu}(2023)}]{hu2023nonhermitianbandtheorydimensions}%
  \BibitemOpen
  \bibfield  {author} {\bibinfo {author} {\bibfnamefont {H.}~\bibnamefont
  {Hu}},\ }\href {https://arxiv.org/abs/2306.12022} {\bibinfo {title}
  {Non-hermitian band theory in all dimensions: uniform spectra and skin
  effect}} (\bibinfo {year} {2023}),\ \Eprint
  {https://arxiv.org/abs/2306.12022} {arXiv:2306.12022 [cond-mat.mes-hall]}
  \BibitemShut {NoStop}%
\bibitem [{\citenamefont {Xiong}\ \emph {et~al.}(2024)\citenamefont {Xiong},
  \citenamefont {Xing},\ and\ \citenamefont
  {Hu}}]{xiong2024nonhermitianskineffectarbitrary}%
  \BibitemOpen
  \bibfield  {author} {\bibinfo {author} {\bibfnamefont {Y.}~\bibnamefont
  {Xiong}}, \bibinfo {author} {\bibfnamefont {Z.-Y.}\ \bibnamefont {Xing}},\
  and\ \bibinfo {author} {\bibfnamefont {H.}~\bibnamefont {Hu}},\ }\href
  {https://arxiv.org/abs/2407.01296} {\bibinfo {title} {Non-hermitian skin
  effect in arbitrary dimensions: non-bloch band theory and classification}}
  (\bibinfo {year} {2024}),\ \Eprint {https://arxiv.org/abs/2407.01296}
  {arXiv:2407.01296 [cond-mat.mes-hall]} \BibitemShut {NoStop}%
\bibitem [{\citenamefont {Hatano}\ and\ \citenamefont
  {Nelson}(1996)}]{hatano1997}%
  \BibitemOpen
  \bibfield  {author} {\bibinfo {author} {\bibfnamefont {N.}~\bibnamefont
  {Hatano}}\ and\ \bibinfo {author} {\bibfnamefont {D.~R.}\ \bibnamefont
  {Nelson}},\ }\bibfield  {title} {\bibinfo {title} {Localization transitions
  in non-hermitian quantum mechanics},\ }\href
  {https://doi.org/10.1103/PhysRevLett.77.570} {\bibfield  {journal} {\bibinfo
  {journal} {Phys. Rev. Lett.}\ }\textbf {\bibinfo {volume} {77}},\ \bibinfo
  {pages} {570} (\bibinfo {year} {1996})}\BibitemShut {NoStop}%
\bibitem [{\citenamefont {Hatano}\ and\ \citenamefont
  {Nelson}(1998)}]{Hatano1998}%
  \BibitemOpen
  \bibfield  {author} {\bibinfo {author} {\bibfnamefont {N.}~\bibnamefont
  {Hatano}}\ and\ \bibinfo {author} {\bibfnamefont {D.~R.}\ \bibnamefont
  {Nelson}},\ }\bibfield  {title} {\bibinfo {title} {Non-hermitian
  delocalization and eigenfunctions},\ }\href
  {https://doi.org/10.1103/PhysRevB.58.8384} {\bibfield  {journal} {\bibinfo
  {journal} {Phys. Rev. B}\ }\textbf {\bibinfo {volume} {58}},\ \bibinfo
  {pages} {8384} (\bibinfo {year} {1998})}\BibitemShut {NoStop}%
\bibitem [{\citenamefont {Longhi}(2019{\natexlab{a}})}]{longhi2019prl}%
  \BibitemOpen
  \bibfield  {author} {\bibinfo {author} {\bibfnamefont {S.}~\bibnamefont
  {Longhi}},\ }\bibfield  {title} {\bibinfo {title} {Topological phase
  transition in non-hermitian quasicrystals},\ }\href
  {https://doi.org/10.1103/PhysRevLett.122.237601} {\bibfield  {journal}
  {\bibinfo  {journal} {Phys. Rev. Lett.}\ }\textbf {\bibinfo {volume} {122}},\
  \bibinfo {pages} {237601} (\bibinfo {year} {2019}{\natexlab{a}})}\BibitemShut
  {NoStop}%
\bibitem [{\citenamefont {Jiang}\ \emph {et~al.}(2019)\citenamefont {Jiang},
  \citenamefont {Lang}, \citenamefont {Yang}, \citenamefont {Zhu},\ and\
  \citenamefont {Chen}}]{hui2019}%
  \BibitemOpen
  \bibfield  {author} {\bibinfo {author} {\bibfnamefont {H.}~\bibnamefont
  {Jiang}}, \bibinfo {author} {\bibfnamefont {L.-J.}\ \bibnamefont {Lang}},
  \bibinfo {author} {\bibfnamefont {C.}~\bibnamefont {Yang}}, \bibinfo {author}
  {\bibfnamefont {S.-L.}\ \bibnamefont {Zhu}},\ and\ \bibinfo {author}
  {\bibfnamefont {S.}~\bibnamefont {Chen}},\ }\bibfield  {title} {\bibinfo
  {title} {Interplay of non-hermitian skin effects and anderson localization in
  nonreciprocal quasiperiodic lattices},\ }\href
  {https://doi.org/10.1103/PhysRevB.100.054301} {\bibfield  {journal} {\bibinfo
   {journal} {Phys. Rev. B}\ }\textbf {\bibinfo {volume} {100}},\ \bibinfo
  {pages} {054301} (\bibinfo {year} {2019})}\BibitemShut {NoStop}%
\bibitem [{\citenamefont {Longhi}(2019{\natexlab{b}})}]{longhi2019}%
  \BibitemOpen
  \bibfield  {author} {\bibinfo {author} {\bibfnamefont {S.}~\bibnamefont
  {Longhi}},\ }\bibfield  {title} {\bibinfo {title} {Metal-insulator phase
  transition in a non-hermitian aubry-andr\'e-harper model},\ }\href
  {https://doi.org/10.1103/PhysRevB.100.125157} {\bibfield  {journal} {\bibinfo
   {journal} {Phys. Rev. B}\ }\textbf {\bibinfo {volume} {100}},\ \bibinfo
  {pages} {125157} (\bibinfo {year} {2019}{\natexlab{b}})}\BibitemShut
  {NoStop}%
\bibitem [{\citenamefont {Liu}\ \emph {et~al.}(2020{\natexlab{a}})\citenamefont
  {Liu}, \citenamefont {Guo}, \citenamefont {Pu},\ and\ \citenamefont
  {Longhi}}]{liu2020general}%
  \BibitemOpen
  \bibfield  {author} {\bibinfo {author} {\bibfnamefont {T.}~\bibnamefont
  {Liu}}, \bibinfo {author} {\bibfnamefont {H.}~\bibnamefont {Guo}}, \bibinfo
  {author} {\bibfnamefont {Y.}~\bibnamefont {Pu}},\ and\ \bibinfo {author}
  {\bibfnamefont {S.}~\bibnamefont {Longhi}},\ }\bibfield  {title} {\bibinfo
  {title} {Generalized aubry-andr\'e self-duality and mobility edges in
  non-hermitian quasiperiodic lattices},\ }\href
  {https://doi.org/10.1103/PhysRevB.102.024205} {\bibfield  {journal} {\bibinfo
   {journal} {Phys. Rev. B}\ }\textbf {\bibinfo {volume} {102}},\ \bibinfo
  {pages} {024205} (\bibinfo {year} {2020}{\natexlab{a}})}\BibitemShut
  {NoStop}%
\bibitem [{\citenamefont {Liu}\ \emph {et~al.}(2020{\natexlab{b}})\citenamefont
  {Liu}, \citenamefont {Jiang}, \citenamefont {Cao},\ and\ \citenamefont
  {Chen}}]{liu2020pt}%
  \BibitemOpen
  \bibfield  {author} {\bibinfo {author} {\bibfnamefont {Y.}~\bibnamefont
  {Liu}}, \bibinfo {author} {\bibfnamefont {X.-P.}\ \bibnamefont {Jiang}},
  \bibinfo {author} {\bibfnamefont {J.}~\bibnamefont {Cao}},\ and\ \bibinfo
  {author} {\bibfnamefont {S.}~\bibnamefont {Chen}},\ }\bibfield  {title}
  {\bibinfo {title} {Non-hermitian mobility edges in one-dimensional
  quasicrystals with parity-time symmetry},\ }\href
  {https://doi.org/10.1103/PhysRevB.101.174205} {\bibfield  {journal} {\bibinfo
   {journal} {Phys. Rev. B}\ }\textbf {\bibinfo {volume} {101}},\ \bibinfo
  {pages} {174205} (\bibinfo {year} {2020}{\natexlab{b}})}\BibitemShut
  {NoStop}%
\bibitem [{\citenamefont {Liu}\ \emph {et~al.}(2021)\citenamefont {Liu},
  \citenamefont {Zhou},\ and\ \citenamefont {Chen}}]{liu2021}%
  \BibitemOpen
  \bibfield  {author} {\bibinfo {author} {\bibfnamefont {Y.}~\bibnamefont
  {Liu}}, \bibinfo {author} {\bibfnamefont {Q.}~\bibnamefont {Zhou}},\ and\
  \bibinfo {author} {\bibfnamefont {S.}~\bibnamefont {Chen}},\ }\bibfield
  {title} {\bibinfo {title} {Localization transition, spectrum structure, and
  winding numbers for one-dimensional non-hermitian quasicrystals},\ }\href
  {https://doi.org/10.1103/PhysRevB.104.024201} {\bibfield  {journal} {\bibinfo
   {journal} {Phys. Rev. B}\ }\textbf {\bibinfo {volume} {104}},\ \bibinfo
  {pages} {024201} (\bibinfo {year} {2021})}\BibitemShut {NoStop}%
\bibitem [{\citenamefont {Yuce}\ and\ \citenamefont
  {Ramezani}(2022)}]{yuce2022}%
  \BibitemOpen
  \bibfield  {author} {\bibinfo {author} {\bibfnamefont {C.}~\bibnamefont
  {Yuce}}\ and\ \bibinfo {author} {\bibfnamefont {H.}~\bibnamefont
  {Ramezani}},\ }\bibfield  {title} {\bibinfo {title} {Coexistence of extended
  and localized states in the one-dimensional non-hermitian anderson model},\
  }\href {https://doi.org/10.1103/PhysRevB.106.024202} {\bibfield  {journal}
  {\bibinfo  {journal} {Phys. Rev. B}\ }\textbf {\bibinfo {volume} {106}},\
  \bibinfo {pages} {024202} (\bibinfo {year} {2022})}\BibitemShut {NoStop}%
\bibitem [{\citenamefont {Zeng}\ \emph {et~al.}(2020)\citenamefont {Zeng},
  \citenamefont {Yang},\ and\ \citenamefont {Xu}}]{zeng2020}%
  \BibitemOpen
  \bibfield  {author} {\bibinfo {author} {\bibfnamefont {Q.-B.}\ \bibnamefont
  {Zeng}}, \bibinfo {author} {\bibfnamefont {Y.-B.}\ \bibnamefont {Yang}},\
  and\ \bibinfo {author} {\bibfnamefont {Y.}~\bibnamefont {Xu}},\ }\bibfield
  {title} {\bibinfo {title} {Topological phases in non-hermitian
  aubry-andr\'e-harper models},\ }\href
  {https://doi.org/10.1103/PhysRevB.101.020201} {\bibfield  {journal} {\bibinfo
   {journal} {Phys. Rev. B}\ }\textbf {\bibinfo {volume} {101}},\ \bibinfo
  {pages} {020201} (\bibinfo {year} {2020})}\BibitemShut {NoStop}%
\bibitem [{\citenamefont {Longhi}(2021{\natexlab{a}})}]{longhi202101}%
  \BibitemOpen
  \bibfield  {author} {\bibinfo {author} {\bibfnamefont {S.}~\bibnamefont
  {Longhi}},\ }\bibfield  {title} {\bibinfo {title} {Phase transitions in a
  non-hermitian aubry-andr\'e-harper model},\ }\href
  {https://doi.org/10.1103/PhysRevB.103.054203} {\bibfield  {journal} {\bibinfo
   {journal} {Phys. Rev. B}\ }\textbf {\bibinfo {volume} {103}},\ \bibinfo
  {pages} {054203} (\bibinfo {year} {2021}{\natexlab{a}})}\BibitemShut
  {NoStop}%
\bibitem [{\citenamefont {Longhi}(2021{\natexlab{b}})}]{longhi202102}%
  \BibitemOpen
  \bibfield  {author} {\bibinfo {author} {\bibfnamefont {S.}~\bibnamefont
  {Longhi}},\ }\bibfield  {title} {\bibinfo {title} {Spectral deformations in
  non-hermitian lattices with disorder and skin effect: A solvable model},\
  }\href {https://doi.org/10.1103/PhysRevB.103.144202} {\bibfield  {journal}
  {\bibinfo  {journal} {Phys. Rev. B}\ }\textbf {\bibinfo {volume} {103}},\
  \bibinfo {pages} {144202} (\bibinfo {year} {2021}{\natexlab{b}})}\BibitemShut
  {NoStop}%
\bibitem [{\citenamefont {Claes}\ and\ \citenamefont
  {Hughes}(2021)}]{claes2021}%
  \BibitemOpen
  \bibfield  {author} {\bibinfo {author} {\bibfnamefont {J.}~\bibnamefont
  {Claes}}\ and\ \bibinfo {author} {\bibfnamefont {T.~L.}\ \bibnamefont
  {Hughes}},\ }\bibfield  {title} {\bibinfo {title} {Skin effect and winding
  number in disordered non-hermitian systems},\ }\href
  {https://doi.org/10.1103/PhysRevB.103.L140201} {\bibfield  {journal}
  {\bibinfo  {journal} {Phys. Rev. B}\ }\textbf {\bibinfo {volume} {103}},\
  \bibinfo {pages} {L140201} (\bibinfo {year} {2021})}\BibitemShut {NoStop}%
\bibitem [{\citenamefont {Okuma}\ and\ \citenamefont
  {Sato}(2021)}]{okuma2021disorder}%
  \BibitemOpen
  \bibfield  {author} {\bibinfo {author} {\bibfnamefont {N.}~\bibnamefont
  {Okuma}}\ and\ \bibinfo {author} {\bibfnamefont {M.}~\bibnamefont {Sato}},\
  }\bibfield  {title} {\bibinfo {title} {Non-hermitian skin effects in
  hermitian correlated or disordered systems: Quantities sensitive or
  insensitive to boundary effects and pseudo-quantum-number},\ }\href
  {https://doi.org/10.1103/PhysRevLett.126.176601} {\bibfield  {journal}
  {\bibinfo  {journal} {Phys. Rev. Lett.}\ }\textbf {\bibinfo {volume} {126}},\
  \bibinfo {pages} {176601} (\bibinfo {year} {2021})}\BibitemShut {NoStop}%
\bibitem [{\citenamefont {Sarkar}\ \emph {et~al.}(2022)\citenamefont {Sarkar},
  \citenamefont {Hegde},\ and\ \citenamefont {Narayan}}]{ronika2022}%
  \BibitemOpen
  \bibfield  {author} {\bibinfo {author} {\bibfnamefont {R.}~\bibnamefont
  {Sarkar}}, \bibinfo {author} {\bibfnamefont {S.~S.}\ \bibnamefont {Hegde}},\
  and\ \bibinfo {author} {\bibfnamefont {A.}~\bibnamefont {Narayan}},\
  }\bibfield  {title} {\bibinfo {title} {Interplay of disorder and point-gap
  topology: Chiral modes, localization, and non-hermitian anderson skin effect
  in one dimension},\ }\href {https://doi.org/10.1103/PhysRevB.106.014207}
  {\bibfield  {journal} {\bibinfo  {journal} {Phys. Rev. B}\ }\textbf {\bibinfo
  {volume} {106}},\ \bibinfo {pages} {014207} (\bibinfo {year}
  {2022})}\BibitemShut {NoStop}%
\bibitem [{\citenamefont {Lin}\ \emph {et~al.}(2022)\citenamefont {Lin},
  \citenamefont {Li}, \citenamefont {Xiao}, \citenamefont {Wang}, \citenamefont
  {Yi},\ and\ \citenamefont {Xue}}]{lin2022obser}%
  \BibitemOpen
  \bibfield  {author} {\bibinfo {author} {\bibfnamefont {Q.}~\bibnamefont
  {Lin}}, \bibinfo {author} {\bibfnamefont {T.}~\bibnamefont {Li}}, \bibinfo
  {author} {\bibfnamefont {L.}~\bibnamefont {Xiao}}, \bibinfo {author}
  {\bibfnamefont {K.}~\bibnamefont {Wang}}, \bibinfo {author} {\bibfnamefont
  {W.}~\bibnamefont {Yi}},\ and\ \bibinfo {author} {\bibfnamefont
  {P.}~\bibnamefont {Xue}},\ }\bibfield  {title} {\bibinfo {title} {Observation
  of non-hermitian topological anderson insulator in quantum dynamics},\ }\href
  {https://doi.org/10.1038/s41467-022-30938-9} {\bibfield  {journal} {\bibinfo
  {journal} {Nature Communications}\ }\textbf {\bibinfo {volume} {13}},\
  \bibinfo {pages} {3229} (\bibinfo {year} {2022})}\BibitemShut {NoStop}%
\bibitem [{\citenamefont {Orito}\ and\ \citenamefont
  {Imura}(2022)}]{orito2022en}%
  \BibitemOpen
  \bibfield  {author} {\bibinfo {author} {\bibfnamefont {T.}~\bibnamefont
  {Orito}}\ and\ \bibinfo {author} {\bibfnamefont {K.-I.}\ \bibnamefont
  {Imura}},\ }\bibfield  {title} {\bibinfo {title} {Unusual wave-packet
  spreading and entanglement dynamics in non-hermitian disordered many-body
  systems},\ }\href {https://doi.org/10.1103/PhysRevB.105.024303} {\bibfield
  {journal} {\bibinfo  {journal} {Phys. Rev. B}\ }\textbf {\bibinfo {volume}
  {105}},\ \bibinfo {pages} {024303} (\bibinfo {year} {2022})}\BibitemShut
  {NoStop}%
\bibitem [{\citenamefont {Kokkinakis}\ \emph {et~al.}(2024)\citenamefont
  {Kokkinakis}, \citenamefont {Makris},\ and\ \citenamefont
  {Economou}}]{kokkinakis2024anderson}%
  \BibitemOpen
  \bibfield  {author} {\bibinfo {author} {\bibfnamefont {E.~T.}\ \bibnamefont
  {Kokkinakis}}, \bibinfo {author} {\bibfnamefont {K.~G.}\ \bibnamefont
  {Makris}},\ and\ \bibinfo {author} {\bibfnamefont {E.~N.}\ \bibnamefont
  {Economou}},\ }\href {https://arxiv.org/abs/2407.10746} {\bibinfo {title}
  {Anderson localization versus hopping asymmetry in a disordered lattice}}
  (\bibinfo {year} {2024}),\ \Eprint {https://arxiv.org/abs/2407.10746}
  {arXiv:2407.10746 [cond-mat.dis-nn]} \BibitemShut {NoStop}%
\bibitem [{\citenamefont {Hamazaki}\ \emph {et~al.}(2019)\citenamefont
  {Hamazaki}, \citenamefont {Kawabata},\ and\ \citenamefont
  {Ueda}}]{PhysRevLett.123.090603}%
  \BibitemOpen
  \bibfield  {author} {\bibinfo {author} {\bibfnamefont {R.}~\bibnamefont
  {Hamazaki}}, \bibinfo {author} {\bibfnamefont {K.}~\bibnamefont {Kawabata}},\
  and\ \bibinfo {author} {\bibfnamefont {M.}~\bibnamefont {Ueda}},\ }\bibfield
  {title} {\bibinfo {title} {Non-hermitian many-body localization},\ }\href
  {https://doi.org/10.1103/PhysRevLett.123.090603} {\bibfield  {journal}
  {\bibinfo  {journal} {Phys. Rev. Lett.}\ }\textbf {\bibinfo {volume} {123}},\
  \bibinfo {pages} {090603} (\bibinfo {year} {2019})}\BibitemShut {NoStop}%
\bibitem [{\citenamefont {Hamazaki}\ \emph {et~al.}(2022)\citenamefont
  {Hamazaki}, \citenamefont {Nakagawa}, \citenamefont {Haga},\ and\
  \citenamefont {Ueda}}]{hamazaki2022lindbladianmanybodylocalization}%
  \BibitemOpen
  \bibfield  {author} {\bibinfo {author} {\bibfnamefont {R.}~\bibnamefont
  {Hamazaki}}, \bibinfo {author} {\bibfnamefont {M.}~\bibnamefont {Nakagawa}},
  \bibinfo {author} {\bibfnamefont {T.}~\bibnamefont {Haga}},\ and\ \bibinfo
  {author} {\bibfnamefont {M.}~\bibnamefont {Ueda}},\ }\href
  {https://arxiv.org/abs/2206.02984} {\bibinfo {title} {Lindbladian many-body
  localization}} (\bibinfo {year} {2022}),\ \Eprint
  {https://arxiv.org/abs/2206.02984} {arXiv:2206.02984 [cond-mat.dis-nn]}
  \BibitemShut {NoStop}%
\bibitem [{\citenamefont {Roccati}\ \emph {et~al.}(2024)\citenamefont
  {Roccati}, \citenamefont {Balducci}, \citenamefont {Shir},\ and\
  \citenamefont {Chenu}}]{PhysRevB.109.L140201}%
  \BibitemOpen
  \bibfield  {author} {\bibinfo {author} {\bibfnamefont {F.}~\bibnamefont
  {Roccati}}, \bibinfo {author} {\bibfnamefont {F.}~\bibnamefont {Balducci}},
  \bibinfo {author} {\bibfnamefont {R.}~\bibnamefont {Shir}},\ and\ \bibinfo
  {author} {\bibfnamefont {A.}~\bibnamefont {Chenu}},\ }\bibfield  {title}
  {\bibinfo {title} {Diagnosing non-hermitian many-body localization and
  quantum chaos via singular value decomposition},\ }\href
  {https://doi.org/10.1103/PhysRevB.109.L140201} {\bibfield  {journal}
  {\bibinfo  {journal} {Phys. Rev. B}\ }\textbf {\bibinfo {volume} {109}},\
  \bibinfo {pages} {L140201} (\bibinfo {year} {2024})}\BibitemShut {NoStop}%
\bibitem [{\citenamefont {De~Tomasi}\ and\ \citenamefont
  {Khaymovich}(2024)}]{PhysRevB.109.174205}%
  \BibitemOpen
  \bibfield  {author} {\bibinfo {author} {\bibfnamefont {G.}~\bibnamefont
  {De~Tomasi}}\ and\ \bibinfo {author} {\bibfnamefont {I.~M.}\ \bibnamefont
  {Khaymovich}},\ }\bibfield  {title} {\bibinfo {title} {Stable many-body
  localization under random continuous measurements in the no-click limit},\
  }\href {https://doi.org/10.1103/PhysRevB.109.174205} {\bibfield  {journal}
  {\bibinfo  {journal} {Phys. Rev. B}\ }\textbf {\bibinfo {volume} {109}},\
  \bibinfo {pages} {174205} (\bibinfo {year} {2024})}\BibitemShut {NoStop}%
\bibitem [{Note1()}]{Note1}%
  \BibitemOpen
  \bibinfo {note} {The case of complex $\mu $ can be simplified to a real $\mu
  $ via a complex phase factor $e^{i\theta }\protect \hat {H}$ upon the
  Hamiltonian and thus a rotation of the complex energy plane, as we elaborate
  in the Supplemental Material \cite {supp}.}\BibitemShut {Stop}%
\bibitem [{\citenamefont {Hu}\ \emph {et~al.}(2023)\citenamefont {Hu},
  \citenamefont {Fu},\ and\ \citenamefont {Zhang}}]{PhysRevB.108.245114}%
  \BibitemOpen
  \bibfield  {author} {\bibinfo {author} {\bibfnamefont {S.-X.}\ \bibnamefont
  {Hu}}, \bibinfo {author} {\bibfnamefont {Y.}~\bibnamefont {Fu}},\ and\
  \bibinfo {author} {\bibfnamefont {Y.}~\bibnamefont {Zhang}},\ }\bibfield
  {title} {\bibinfo {title} {Nontrivial worldline winding in non-hermitian
  quantum systems},\ }\href {https://doi.org/10.1103/PhysRevB.108.245114}
  {\bibfield  {journal} {\bibinfo  {journal} {Phys. Rev. B}\ }\textbf {\bibinfo
  {volume} {108}},\ \bibinfo {pages} {245114} (\bibinfo {year}
  {2023})}\BibitemShut {NoStop}%
\bibitem [{Note2()}]{Note2}%
  \BibitemOpen
  \bibinfo {note} {Admittedly, the microscopic mechanism that an open or
  non-equilibrium system may induce residue $\protect \operatorname {Im}(v)$
  like a non-Hermitian Hamiltonian remains an open question for future
  research.}\BibitemShut {Stop}%
\bibitem [{sup()}]{supp}%
  \BibitemOpen
  \href@noop {} {\bibinfo {title} {Please refer to the supplemental material
  for further details.}}\BibitemShut {Stop}%
\bibitem [{Note3()}]{Note3}%
  \BibitemOpen
  \bibinfo {note} {Multiple bands or Fermi Seas, if present, also need to be
  summed over.}\BibitemShut {Stop}%
\bibitem [{Note4()}]{Note4}%
  \BibitemOpen
  \bibinfo {note} {It also implies that a residue $\protect \operatorname
  {Im}(v)$ under PBC is related to the NHSE under OBC.}\BibitemShut {Stop}%
\bibitem [{Note5()}]{Note5}%
  \BibitemOpen
  \bibinfo {note} {Similarly to Hermitian quantum systems, localization begins
  at the band edges and moves toward the center.}\BibitemShut {Stop}%
\bibitem [{\citenamefont {Liang}\ \emph {et~al.}(2022)\citenamefont {Liang},
  \citenamefont {Xie}, \citenamefont {Dong}, \citenamefont {Li}, \citenamefont
  {Li}, \citenamefont {Gadway}, \citenamefont {Yi},\ and\ \citenamefont
  {Yan}}]{triangleNHmodel}%
  \BibitemOpen
  \bibfield  {author} {\bibinfo {author} {\bibfnamefont {Q.}~\bibnamefont
  {Liang}}, \bibinfo {author} {\bibfnamefont {D.}~\bibnamefont {Xie}}, \bibinfo
  {author} {\bibfnamefont {Z.}~\bibnamefont {Dong}}, \bibinfo {author}
  {\bibfnamefont {H.}~\bibnamefont {Li}}, \bibinfo {author} {\bibfnamefont
  {H.}~\bibnamefont {Li}}, \bibinfo {author} {\bibfnamefont {B.}~\bibnamefont
  {Gadway}}, \bibinfo {author} {\bibfnamefont {W.}~\bibnamefont {Yi}},\ and\
  \bibinfo {author} {\bibfnamefont {B.}~\bibnamefont {Yan}},\ }\bibfield
  {title} {\bibinfo {title} {Dynamic signatures of non-hermitian skin effect
  and topology in ultracold atoms},\ }\href
  {https://doi.org/10.1103/PhysRevLett.129.070401} {\bibfield  {journal}
  {\bibinfo  {journal} {Phys. Rev. Lett.}\ }\textbf {\bibinfo {volume} {129}},\
  \bibinfo {pages} {070401} (\bibinfo {year} {2022})}\BibitemShut {NoStop}%
\bibitem [{\citenamefont {Wang}\ \emph {et~al.}(2025)\citenamefont {Wang},
  \citenamefont {Wang},\ and\ \citenamefont {Ma}}]{Ma2025}%
  \BibitemOpen
  \bibfield  {author} {\bibinfo {author} {\bibfnamefont {W.}~\bibnamefont
  {Wang}}, \bibinfo {author} {\bibfnamefont {X.}~\bibnamefont {Wang}},\ and\
  \bibinfo {author} {\bibfnamefont {G.}~\bibnamefont {Ma}},\ }\bibfield
  {title} {\bibinfo {title} {Anderson transition at complex energies in
  one-dimensional parity-time-symmetric disordered systems},\ }\href
  {https://doi.org/10.1103/PhysRevLett.134.066301} {\bibfield  {journal}
  {\bibinfo  {journal} {Phys. Rev. Lett.}\ }\textbf {\bibinfo {volume} {134}},\
  \bibinfo {pages} {066301} (\bibinfo {year} {2025})}\BibitemShut {NoStop}%
\bibitem [{Note6()}]{Note6}%
  \BibitemOpen
  \bibinfo {note} {On the contrary, localized physical properties require both
  localized single-particle states and $\protect \operatorname {Im}(v) =
  0$.}\BibitemShut {Stop}%
\bibitem [{\citenamefont {Echeverri-Arteaga}\ \emph {et~al.}(2018)\citenamefont
  {Echeverri-Arteaga}, \citenamefont {Vinck-Posada},\ and\ \citenamefont
  {Gómez}}]{ECHEVERRIARTEAGA2018413}%
  \BibitemOpen
  \bibfield  {author} {\bibinfo {author} {\bibfnamefont {S.}~\bibnamefont
  {Echeverri-Arteaga}}, \bibinfo {author} {\bibfnamefont {H.}~\bibnamefont
  {Vinck-Posada}},\ and\ \bibinfo {author} {\bibfnamefont {E.~A.}\ \bibnamefont
  {Gómez}},\ }\bibfield  {title} {\bibinfo {title} {A comparative study on the
  reliability of non-hermitian effective hamiltonian approach for modeling open
  quantum systems},\ }\href
  {https://doi.org/https://doi.org/10.1016/j.ijleo.2018.06.081} {\bibfield
  {journal} {\bibinfo  {journal} {Optik}\ }\textbf {\bibinfo {volume} {171}},\
  \bibinfo {pages} {413} (\bibinfo {year} {2018})}\BibitemShut {NoStop}%
\bibitem [{\citenamefont {Reisenbauer}\ \emph {et~al.}(2024)\citenamefont
  {Reisenbauer}, \citenamefont {Rudolph}, \citenamefont {Egyed}, \citenamefont
  {Hornberger}, \citenamefont {Zasedatelev}, \citenamefont {Abuzarli},
  \citenamefont {Stickler},\ and\ \citenamefont
  {Deli{\'{c}}}}]{Reisenbauer2024}%
  \BibitemOpen
  \bibfield  {author} {\bibinfo {author} {\bibfnamefont {M.}~\bibnamefont
  {Reisenbauer}}, \bibinfo {author} {\bibfnamefont {H.}~\bibnamefont
  {Rudolph}}, \bibinfo {author} {\bibfnamefont {L.}~\bibnamefont {Egyed}},
  \bibinfo {author} {\bibfnamefont {K.}~\bibnamefont {Hornberger}}, \bibinfo
  {author} {\bibfnamefont {A.~V.}\ \bibnamefont {Zasedatelev}}, \bibinfo
  {author} {\bibfnamefont {M.}~\bibnamefont {Abuzarli}}, \bibinfo {author}
  {\bibfnamefont {B.~A.}\ \bibnamefont {Stickler}},\ and\ \bibinfo {author}
  {\bibfnamefont {U.}~\bibnamefont {Deli{\'{c}}}},\ }\bibfield  {title}
  {\bibinfo {title} {Non-hermitian dynamics and non-reciprocity of optically
  coupled nanoparticles},\ }\href {https://doi.org/10.1038/s41567-024-02589-8}
  {\bibfield  {journal} {\bibinfo  {journal} {Nature Physics}\ }\textbf
  {\bibinfo {volume} {20}},\ \bibinfo {pages} {1629} (\bibinfo {year}
  {2024})}\BibitemShut {NoStop}%
\bibitem [{\citenamefont {Orito}\ and\ \citenamefont
  {Imura}(2023)}]{PhysRevB.108.214308}%
  \BibitemOpen
  \bibfield  {author} {\bibinfo {author} {\bibfnamefont {T.}~\bibnamefont
  {Orito}}\ and\ \bibinfo {author} {\bibfnamefont {K.-I.}\ \bibnamefont
  {Imura}},\ }\bibfield  {title} {\bibinfo {title} {Entanglement dynamics in
  the many-body hatano-nelson model},\ }\href
  {https://doi.org/10.1103/PhysRevB.108.214308} {\bibfield  {journal} {\bibinfo
   {journal} {Phys. Rev. B}\ }\textbf {\bibinfo {volume} {108}},\ \bibinfo
  {pages} {214308} (\bibinfo {year} {2023})}\BibitemShut {NoStop}%
\bibitem [{\citenamefont {Wang}\ \emph
  {et~al.}(2024{\natexlab{b}})\citenamefont {Wang}, \citenamefont {Fang},\ and\
  \citenamefont {Ren}}]{PhysRevB.110.035113}%
  \BibitemOpen
  \bibfield  {author} {\bibinfo {author} {\bibfnamefont {Y.-P.}\ \bibnamefont
  {Wang}}, \bibinfo {author} {\bibfnamefont {C.}~\bibnamefont {Fang}},\ and\
  \bibinfo {author} {\bibfnamefont {J.}~\bibnamefont {Ren}},\ }\bibfield
  {title} {\bibinfo {title} {Absence of measurement-induced entanglement
  transition due to feedback-induced skin effect},\ }\href
  {https://doi.org/10.1103/PhysRevB.110.035113} {\bibfield  {journal} {\bibinfo
   {journal} {Phys. Rev. B}\ }\textbf {\bibinfo {volume} {110}},\ \bibinfo
  {pages} {035113} (\bibinfo {year} {2024}{\natexlab{b}})}\BibitemShut
  {NoStop}%
\bibitem [{\citenamefont {Zhang}\ \emph {et~al.}(2025)\citenamefont {Zhang},
  \citenamefont {Carrasquilla},\ and\ \citenamefont {Kim}}]{ZhangYuxuan2025}%
  \BibitemOpen
  \bibfield  {author} {\bibinfo {author} {\bibfnamefont {Y.}~\bibnamefont
  {Zhang}}, \bibinfo {author} {\bibfnamefont {J.}~\bibnamefont
  {Carrasquilla}},\ and\ \bibinfo {author} {\bibfnamefont {Y.~B.}\ \bibnamefont
  {Kim}},\ }\bibfield  {title} {\bibinfo {title} {Observation of a
  non-hermitian supersonic mode on a trapped-ion quantum computer},\ }\href
  {https://doi.org/10.1038/s41467-025-57930-3} {\bibfield  {journal} {\bibinfo
  {journal} {Nature Communications}\ }\textbf {\bibinfo {volume} {16}},\
  \bibinfo {pages} {3286} (\bibinfo {year} {2025})}\BibitemShut {NoStop}%
\bibitem [{\citenamefont {Gou}\ \emph {et~al.}(2020)\citenamefont {Gou},
  \citenamefont {Chen}, \citenamefont {Xie}, \citenamefont {Xiao},
  \citenamefont {Deng}, \citenamefont {Gadway}, \citenamefont {Yi},\ and\
  \citenamefont {Yan}}]{Boyan2020}%
  \BibitemOpen
  \bibfield  {author} {\bibinfo {author} {\bibfnamefont {W.}~\bibnamefont
  {Gou}}, \bibinfo {author} {\bibfnamefont {T.}~\bibnamefont {Chen}}, \bibinfo
  {author} {\bibfnamefont {D.}~\bibnamefont {Xie}}, \bibinfo {author}
  {\bibfnamefont {T.}~\bibnamefont {Xiao}}, \bibinfo {author} {\bibfnamefont
  {T.-S.}\ \bibnamefont {Deng}}, \bibinfo {author} {\bibfnamefont
  {B.}~\bibnamefont {Gadway}}, \bibinfo {author} {\bibfnamefont
  {W.}~\bibnamefont {Yi}},\ and\ \bibinfo {author} {\bibfnamefont
  {B.}~\bibnamefont {Yan}},\ }\bibfield  {title} {\bibinfo {title} {Tunable
  nonreciprocal quantum transport through a dissipative aharonov-bohm ring in
  ultracold atoms},\ }\href {https://doi.org/10.1103/PhysRevLett.124.070402}
  {\bibfield  {journal} {\bibinfo  {journal} {Phys. Rev. Lett.}\ }\textbf
  {\bibinfo {volume} {124}},\ \bibinfo {pages} {070402} (\bibinfo {year}
  {2020})}\BibitemShut {NoStop}%
\bibitem [{\citenamefont {Lapp}\ \emph {et~al.}(2019)\citenamefont {Lapp},
  \citenamefont {Ang’ong’a}, \citenamefont {An},\ and\ \citenamefont
  {Gadway}}]{Lapp2019}%
  \BibitemOpen
  \bibfield  {author} {\bibinfo {author} {\bibfnamefont {S.}~\bibnamefont
  {Lapp}}, \bibinfo {author} {\bibfnamefont {J.}~\bibnamefont {Ang’ong’a}},
  \bibinfo {author} {\bibfnamefont {F.~A.}\ \bibnamefont {An}},\ and\ \bibinfo
  {author} {\bibfnamefont {B.}~\bibnamefont {Gadway}},\ }\bibfield  {title}
  {\bibinfo {title} {Engineering tunable local loss in a synthetic lattice of
  momentum states},\ }\href {https://doi.org/10.1088/1367-2630/ab1147}
  {\bibfield  {journal} {\bibinfo  {journal} {New Journal of Physics}\ }\textbf
  {\bibinfo {volume} {21}},\ \bibinfo {pages} {045006} (\bibinfo {year}
  {2019})}\BibitemShut {NoStop}%
\bibitem [{\citenamefont {Xiao}\ \emph {et~al.}(2024)\citenamefont {Xiao},
  \citenamefont {Xue}, \citenamefont {Song}, \citenamefont {Hu}, \citenamefont
  {Yi}, \citenamefont {Wang},\ and\ \citenamefont {Xue}}]{Xue2024}%
  \BibitemOpen
  \bibfield  {author} {\bibinfo {author} {\bibfnamefont {L.}~\bibnamefont
  {Xiao}}, \bibinfo {author} {\bibfnamefont {W.-T.}\ \bibnamefont {Xue}},
  \bibinfo {author} {\bibfnamefont {F.}~\bibnamefont {Song}}, \bibinfo {author}
  {\bibfnamefont {Y.-M.}\ \bibnamefont {Hu}}, \bibinfo {author} {\bibfnamefont
  {W.}~\bibnamefont {Yi}}, \bibinfo {author} {\bibfnamefont {Z.}~\bibnamefont
  {Wang}},\ and\ \bibinfo {author} {\bibfnamefont {P.}~\bibnamefont {Xue}},\
  }\bibfield  {title} {\bibinfo {title} {Observation of non-hermitian edge
  burst in quantum dynamics},\ }\href
  {https://doi.org/10.1103/PhysRevLett.133.070801} {\bibfield  {journal}
  {\bibinfo  {journal} {Phys. Rev. Lett.}\ }\textbf {\bibinfo {volume} {133}},\
  \bibinfo {pages} {070801} (\bibinfo {year} {2024})}\BibitemShut {NoStop}%
\bibitem [{\citenamefont {Gu}\ \emph {et~al.}(2021)\citenamefont {Gu},
  \citenamefont {Gao}, \citenamefont {Cao}, \citenamefont {Liu}, \citenamefont
  {Zhu},\ and\ \citenamefont {Zhu}}]{gu2021controlling}%
  \BibitemOpen
  \bibfield  {author} {\bibinfo {author} {\bibfnamefont {Z.}~\bibnamefont
  {Gu}}, \bibinfo {author} {\bibfnamefont {H.}~\bibnamefont {Gao}}, \bibinfo
  {author} {\bibfnamefont {P.-C.}\ \bibnamefont {Cao}}, \bibinfo {author}
  {\bibfnamefont {T.}~\bibnamefont {Liu}}, \bibinfo {author} {\bibfnamefont
  {X.-F.}\ \bibnamefont {Zhu}},\ and\ \bibinfo {author} {\bibfnamefont
  {J.}~\bibnamefont {Zhu}},\ }\bibfield  {title} {\bibinfo {title} {Controlling
  sound in non-hermitian acoustic systems},\ }\href
  {https://doi.org/10.1103/PhysRevApplied.16.057001} {\bibfield  {journal}
  {\bibinfo  {journal} {Phys. Rev. Appl.}\ }\textbf {\bibinfo {volume} {16}},\
  \bibinfo {pages} {057001} (\bibinfo {year} {2021})}\BibitemShut {NoStop}%
\bibitem [{\citenamefont {Zhang}\ \emph {et~al.}(2021)\citenamefont {Zhang},
  \citenamefont {Yang}, \citenamefont {Ge}, \citenamefont {Guan}, \citenamefont
  {Chen}, \citenamefont {Yan}, \citenamefont {Chen}, \citenamefont {Xi},
  \citenamefont {Li}, \citenamefont {Jia}, \citenamefont {Yuan}, \citenamefont
  {Sun}, \citenamefont {Chen},\ and\ \citenamefont {Zhang}}]{Zhang2021}%
  \BibitemOpen
  \bibfield  {author} {\bibinfo {author} {\bibfnamefont {L.}~\bibnamefont
  {Zhang}}, \bibinfo {author} {\bibfnamefont {Y.}~\bibnamefont {Yang}},
  \bibinfo {author} {\bibfnamefont {Y.}~\bibnamefont {Ge}}, \bibinfo {author}
  {\bibfnamefont {Y.-J.}\ \bibnamefont {Guan}}, \bibinfo {author}
  {\bibfnamefont {Q.}~\bibnamefont {Chen}}, \bibinfo {author} {\bibfnamefont
  {Q.}~\bibnamefont {Yan}}, \bibinfo {author} {\bibfnamefont {F.}~\bibnamefont
  {Chen}}, \bibinfo {author} {\bibfnamefont {R.}~\bibnamefont {Xi}}, \bibinfo
  {author} {\bibfnamefont {Y.}~\bibnamefont {Li}}, \bibinfo {author}
  {\bibfnamefont {D.}~\bibnamefont {Jia}}, \bibinfo {author} {\bibfnamefont
  {S.-Q.}\ \bibnamefont {Yuan}}, \bibinfo {author} {\bibfnamefont {H.-X.}\
  \bibnamefont {Sun}}, \bibinfo {author} {\bibfnamefont {H.}~\bibnamefont
  {Chen}},\ and\ \bibinfo {author} {\bibfnamefont {B.}~\bibnamefont {Zhang}},\
  }\bibfield  {title} {\bibinfo {title} {Acoustic non-hermitian skin effect
  from twisted winding topology},\ }\href
  {https://doi.org/10.1038/s41467-021-26619-8} {\bibfield  {journal} {\bibinfo
  {journal} {Nature Communications}\ }\textbf {\bibinfo {volume} {12}},\
  \bibinfo {pages} {6297} (\bibinfo {year} {2021})}\BibitemShut {NoStop}%
\bibitem [{\citenamefont {Wen}\ \emph {et~al.}(2022)\citenamefont {Wen},
  \citenamefont {Zhu}, \citenamefont {Fan}, \citenamefont {Tam}, \citenamefont
  {Zhu}, \citenamefont {Wu}, \citenamefont {Lemoult}, \citenamefont {Fink},\
  and\ \citenamefont {Li}}]{Wen2022}%
  \BibitemOpen
  \bibfield  {author} {\bibinfo {author} {\bibfnamefont {X.}~\bibnamefont
  {Wen}}, \bibinfo {author} {\bibfnamefont {X.}~\bibnamefont {Zhu}}, \bibinfo
  {author} {\bibfnamefont {A.}~\bibnamefont {Fan}}, \bibinfo {author}
  {\bibfnamefont {W.~Y.}\ \bibnamefont {Tam}}, \bibinfo {author} {\bibfnamefont
  {J.}~\bibnamefont {Zhu}}, \bibinfo {author} {\bibfnamefont {H.~W.}\
  \bibnamefont {Wu}}, \bibinfo {author} {\bibfnamefont {F.}~\bibnamefont
  {Lemoult}}, \bibinfo {author} {\bibfnamefont {M.}~\bibnamefont {Fink}},\ and\
  \bibinfo {author} {\bibfnamefont {J.}~\bibnamefont {Li}},\ }\bibfield
  {title} {\bibinfo {title} {Unidirectional amplification with acoustic
  non-hermitian space-time varying metamaterial},\ }\href
  {https://doi.org/10.1038/s42005-021-00790-2} {\bibfield  {journal} {\bibinfo
  {journal} {Communications Physics}\ }\textbf {\bibinfo {volume} {5}},\
  \bibinfo {pages} {18} (\bibinfo {year} {2022})}\BibitemShut {NoStop}%
\bibitem [{\citenamefont {Fan}\ \emph {et~al.}(2023)\citenamefont {Fan},
  \citenamefont {Gao}, \citenamefont {Liu}, \citenamefont {An}, \citenamefont
  {Kong}, \citenamefont {Xu}, \citenamefont {Zhu}, \citenamefont {Qiu},\ and\
  \citenamefont {Su}}]{fan2023reconf}%
  \BibitemOpen
  \bibfield  {author} {\bibinfo {author} {\bibfnamefont {H.}~\bibnamefont
  {Fan}}, \bibinfo {author} {\bibfnamefont {H.}~\bibnamefont {Gao}}, \bibinfo
  {author} {\bibfnamefont {T.}~\bibnamefont {Liu}}, \bibinfo {author}
  {\bibfnamefont {S.}~\bibnamefont {An}}, \bibinfo {author} {\bibfnamefont
  {X.}~\bibnamefont {Kong}}, \bibinfo {author} {\bibfnamefont {G.}~\bibnamefont
  {Xu}}, \bibinfo {author} {\bibfnamefont {J.}~\bibnamefont {Zhu}}, \bibinfo
  {author} {\bibfnamefont {C.-W.}\ \bibnamefont {Qiu}},\ and\ \bibinfo {author}
  {\bibfnamefont {Z.}~\bibnamefont {Su}},\ }\bibfield  {title} {\bibinfo
  {title} {Reconfigurable topological modes in acoustic non-hermitian
  crystals},\ }\href {https://doi.org/10.1103/PhysRevB.107.L201108} {\bibfield
  {journal} {\bibinfo  {journal} {Phys. Rev. B}\ }\textbf {\bibinfo {volume}
  {107}},\ \bibinfo {pages} {L201108} (\bibinfo {year} {2023})}\BibitemShut
  {NoStop}%
\bibitem [{\citenamefont {Huang}\ \emph {et~al.}(2024)\citenamefont {Huang},
  \citenamefont {Huang}, \citenamefont {Shen}, \citenamefont {Yves},
  \citenamefont {Pilipchuk}, \citenamefont {Ni}, \citenamefont {Kim},
  \citenamefont {Chiang}, \citenamefont {Powell}, \citenamefont {Zhu},
  \citenamefont {Cheng}, \citenamefont {Li}, \citenamefont {Sadreev},
  \citenamefont {Al{\`u}},\ and\ \citenamefont {Miroshnichenko}}]{Huang2024}%
  \BibitemOpen
  \bibfield  {author} {\bibinfo {author} {\bibfnamefont {L.}~\bibnamefont
  {Huang}}, \bibinfo {author} {\bibfnamefont {S.}~\bibnamefont {Huang}},
  \bibinfo {author} {\bibfnamefont {C.}~\bibnamefont {Shen}}, \bibinfo {author}
  {\bibfnamefont {S.}~\bibnamefont {Yves}}, \bibinfo {author} {\bibfnamefont
  {A.~S.}\ \bibnamefont {Pilipchuk}}, \bibinfo {author} {\bibfnamefont
  {X.}~\bibnamefont {Ni}}, \bibinfo {author} {\bibfnamefont {S.}~\bibnamefont
  {Kim}}, \bibinfo {author} {\bibfnamefont {Y.~K.}\ \bibnamefont {Chiang}},
  \bibinfo {author} {\bibfnamefont {D.~A.}\ \bibnamefont {Powell}}, \bibinfo
  {author} {\bibfnamefont {J.}~\bibnamefont {Zhu}}, \bibinfo {author}
  {\bibfnamefont {Y.}~\bibnamefont {Cheng}}, \bibinfo {author} {\bibfnamefont
  {Y.}~\bibnamefont {Li}}, \bibinfo {author} {\bibfnamefont {A.~F.}\
  \bibnamefont {Sadreev}}, \bibinfo {author} {\bibfnamefont {A.}~\bibnamefont
  {Al{\`u}}},\ and\ \bibinfo {author} {\bibfnamefont {A.~E.}\ \bibnamefont
  {Miroshnichenko}},\ }\bibfield  {title} {\bibinfo {title} {Acoustic
  resonances in non-hermitian open systems},\ }\href
  {https://doi.org/10.1038/s42254-023-00659-z} {\bibfield  {journal} {\bibinfo
  {journal} {Nature Reviews Physics}\ }\textbf {\bibinfo {volume} {6}},\
  \bibinfo {pages} {11} (\bibinfo {year} {2024})}\BibitemShut {NoStop}%
\bibitem [{\citenamefont {Brandenbourger}\ \emph {et~al.}(2019)\citenamefont
  {Brandenbourger}, \citenamefont {Locsin}, \citenamefont {Lerner},\ and\
  \citenamefont {Coulais}}]{brandenbourger2019}%
  \BibitemOpen
  \bibfield  {author} {\bibinfo {author} {\bibfnamefont {M.}~\bibnamefont
  {Brandenbourger}}, \bibinfo {author} {\bibfnamefont {X.}~\bibnamefont
  {Locsin}}, \bibinfo {author} {\bibfnamefont {E.}~\bibnamefont {Lerner}},\
  and\ \bibinfo {author} {\bibfnamefont {C.}~\bibnamefont {Coulais}},\
  }\bibfield  {title} {\bibinfo {title} {Non-reciprocal robotic
  metamaterials},\ }\href {https://doi.org/10.1038/s41467-019-12599-3}
  {\bibfield  {journal} {\bibinfo  {journal} {Nature Communications}\ }\textbf
  {\bibinfo {volume} {10}},\ \bibinfo {pages} {4608} (\bibinfo {year}
  {2019})}\BibitemShut {NoStop}%
\bibitem [{\citenamefont {Ghatak}\ \emph {et~al.}(2020)\citenamefont {Ghatak},
  \citenamefont {Brandenbourger}, \citenamefont {van Wezel},\ and\
  \citenamefont {Coulais}}]{ananya2020}%
  \BibitemOpen
  \bibfield  {author} {\bibinfo {author} {\bibfnamefont {A.}~\bibnamefont
  {Ghatak}}, \bibinfo {author} {\bibfnamefont {M.}~\bibnamefont
  {Brandenbourger}}, \bibinfo {author} {\bibfnamefont {J.}~\bibnamefont {van
  Wezel}},\ and\ \bibinfo {author} {\bibfnamefont {C.}~\bibnamefont
  {Coulais}},\ }\bibfield  {title} {\bibinfo {title} {Observation of
  non-hermitian topology and its bulk–edge correspondence in an active
  mechanical metamaterial},\ }\href {https://doi.org/10.1073/pnas.2010580117}
  {\bibfield  {journal} {\bibinfo  {journal} {Proceedings of the National
  Academy of Sciences}\ }\textbf {\bibinfo {volume} {117}},\ \bibinfo {pages}
  {29561} (\bibinfo {year} {2020})}\BibitemShut {NoStop}%
\bibitem [{\citenamefont {Chen}\ \emph {et~al.}(2021)\citenamefont {Chen},
  \citenamefont {Li}, \citenamefont {Scheibner}, \citenamefont {Vitelli},\ and\
  \citenamefont {Huang}}]{chen2021}%
  \BibitemOpen
  \bibfield  {author} {\bibinfo {author} {\bibfnamefont {Y.}~\bibnamefont
  {Chen}}, \bibinfo {author} {\bibfnamefont {X.}~\bibnamefont {Li}}, \bibinfo
  {author} {\bibfnamefont {C.}~\bibnamefont {Scheibner}}, \bibinfo {author}
  {\bibfnamefont {V.}~\bibnamefont {Vitelli}},\ and\ \bibinfo {author}
  {\bibfnamefont {G.}~\bibnamefont {Huang}},\ }\bibfield  {title} {\bibinfo
  {title} {Realization of active metamaterials with odd micropolar
  elasticity},\ }\href {https://doi.org/10.1038/s41467-021-26034-z} {\bibfield
  {journal} {\bibinfo  {journal} {Nature Communications}\ }\textbf {\bibinfo
  {volume} {12}},\ \bibinfo {pages} {5935} (\bibinfo {year}
  {2021})}\BibitemShut {NoStop}%
\bibitem [{\citenamefont {Wang}\ \emph {et~al.}(2022)\citenamefont {Wang},
  \citenamefont {Wang},\ and\ \citenamefont {Ma}}]{wang2022morphing}%
  \BibitemOpen
  \bibfield  {author} {\bibinfo {author} {\bibfnamefont {W.}~\bibnamefont
  {Wang}}, \bibinfo {author} {\bibfnamefont {X.}~\bibnamefont {Wang}},\ and\
  \bibinfo {author} {\bibfnamefont {G.}~\bibnamefont {Ma}},\ }\bibfield
  {title} {\bibinfo {title} {Non-hermitian morphing of topological modes},\
  }\href {https://doi.org/10.1038/s41586-022-04929-1} {\bibfield  {journal}
  {\bibinfo  {journal} {Nature}\ }\textbf {\bibinfo {volume} {608}},\ \bibinfo
  {pages} {50} (\bibinfo {year} {2022})}\BibitemShut {NoStop}%
\bibitem [{\citenamefont {Wang}\ \emph {et~al.}(2023)\citenamefont {Wang},
  \citenamefont {Hu}, \citenamefont {Wang}, \citenamefont {Ma},\ and\
  \citenamefont {Ding}}]{wang2023exp}%
  \BibitemOpen
  \bibfield  {author} {\bibinfo {author} {\bibfnamefont {W.}~\bibnamefont
  {Wang}}, \bibinfo {author} {\bibfnamefont {M.}~\bibnamefont {Hu}}, \bibinfo
  {author} {\bibfnamefont {X.}~\bibnamefont {Wang}}, \bibinfo {author}
  {\bibfnamefont {G.}~\bibnamefont {Ma}},\ and\ \bibinfo {author}
  {\bibfnamefont {K.}~\bibnamefont {Ding}},\ }\bibfield  {title} {\bibinfo
  {title} {Experimental realization of geometry-dependent skin effect in a
  reciprocal two-dimensional lattice},\ }\href
  {https://doi.org/10.1103/PhysRevLett.131.207201} {\bibfield  {journal}
  {\bibinfo  {journal} {Phys. Rev. Lett.}\ }\textbf {\bibinfo {volume} {131}},\
  \bibinfo {pages} {207201} (\bibinfo {year} {2023})}\BibitemShut {NoStop}%
\bibitem [{\citenamefont {Li}\ \emph {et~al.}(2024)\citenamefont {Li},
  \citenamefont {Wang}, \citenamefont {Wang}, \citenamefont {Lin},
  \citenamefont {Ma},\ and\ \citenamefont {Jiang}}]{li2024obser}%
  \BibitemOpen
  \bibfield  {author} {\bibinfo {author} {\bibfnamefont {Z.}~\bibnamefont
  {Li}}, \bibinfo {author} {\bibfnamefont {L.-W.}\ \bibnamefont {Wang}},
  \bibinfo {author} {\bibfnamefont {X.}~\bibnamefont {Wang}}, \bibinfo {author}
  {\bibfnamefont {Z.-K.}\ \bibnamefont {Lin}}, \bibinfo {author} {\bibfnamefont
  {G.}~\bibnamefont {Ma}},\ and\ \bibinfo {author} {\bibfnamefont {J.-H.}\
  \bibnamefont {Jiang}},\ }\bibfield  {title} {\bibinfo {title} {Observation of
  dynamic non-hermitian skin effects},\ }\href
  {https://doi.org/10.1038/s41467-024-50776-1} {\bibfield  {journal} {\bibinfo
  {journal} {Nature Communications}\ }\textbf {\bibinfo {volume} {15}},\
  \bibinfo {pages} {6544} (\bibinfo {year} {2024})}\BibitemShut {NoStop}%
\bibitem [{\citenamefont {Li}\ \emph {et~al.}(2020)\citenamefont {Li},
  \citenamefont {Dong}, \citenamefont {Wang}, \citenamefont {Cheng},
  \citenamefont {Ho}, \citenamefont {Zhang}, \citenamefont {Wen}, \citenamefont
  {Zhang}, \citenamefont {Chan}, \citenamefont {Al\`u}, \citenamefont {Qiu},\
  and\ \citenamefont {Chen}}]{PhysRevLett.125.187403}%
  \BibitemOpen
  \bibfield  {author} {\bibinfo {author} {\bibfnamefont {A.}~\bibnamefont
  {Li}}, \bibinfo {author} {\bibfnamefont {J.}~\bibnamefont {Dong}}, \bibinfo
  {author} {\bibfnamefont {J.}~\bibnamefont {Wang}}, \bibinfo {author}
  {\bibfnamefont {Z.}~\bibnamefont {Cheng}}, \bibinfo {author} {\bibfnamefont
  {J.~S.}\ \bibnamefont {Ho}}, \bibinfo {author} {\bibfnamefont
  {D.}~\bibnamefont {Zhang}}, \bibinfo {author} {\bibfnamefont
  {J.}~\bibnamefont {Wen}}, \bibinfo {author} {\bibfnamefont {X.-L.}\
  \bibnamefont {Zhang}}, \bibinfo {author} {\bibfnamefont {C.~T.}\ \bibnamefont
  {Chan}}, \bibinfo {author} {\bibfnamefont {A.}~\bibnamefont {Al\`u}},
  \bibinfo {author} {\bibfnamefont {C.-W.}\ \bibnamefont {Qiu}},\ and\ \bibinfo
  {author} {\bibfnamefont {L.}~\bibnamefont {Chen}},\ }\bibfield  {title}
  {\bibinfo {title} {Hamiltonian hopping for efficient chiral mode switching in
  encircling exceptional points},\ }\href
  {https://doi.org/10.1103/PhysRevLett.125.187403} {\bibfield  {journal}
  {\bibinfo  {journal} {Phys. Rev. Lett.}\ }\textbf {\bibinfo {volume} {125}},\
  \bibinfo {pages} {187403} (\bibinfo {year} {2020})}\BibitemShut {NoStop}%
\bibitem [{\citenamefont {Maddi}\ \emph {et~al.}(2024)\citenamefont {Maddi},
  \citenamefont {Auregan}, \citenamefont {Penelet}, \citenamefont {Pagneux},\
  and\ \citenamefont {Achilleos}}]{PhysRevResearch.6.L012061}%
  \BibitemOpen
  \bibfield  {author} {\bibinfo {author} {\bibfnamefont {A.}~\bibnamefont
  {Maddi}}, \bibinfo {author} {\bibfnamefont {Y.}~\bibnamefont {Auregan}},
  \bibinfo {author} {\bibfnamefont {G.}~\bibnamefont {Penelet}}, \bibinfo
  {author} {\bibfnamefont {V.}~\bibnamefont {Pagneux}},\ and\ \bibinfo {author}
  {\bibfnamefont {V.}~\bibnamefont {Achilleos}},\ }\bibfield  {title} {\bibinfo
  {title} {Exact analog of the hatano-nelson model in one-dimensional
  continuous nonreciprocal systems},\ }\href
  {https://doi.org/10.1103/PhysRevResearch.6.L012061} {\bibfield  {journal}
  {\bibinfo  {journal} {Phys. Rev. Res.}\ }\textbf {\bibinfo {volume} {6}},\
  \bibinfo {pages} {L012061} (\bibinfo {year} {2024})}\BibitemShut {NoStop}%
\bibitem [{\citenamefont {Geng}\ \emph {et~al.}(2023)\citenamefont {Geng},
  \citenamefont {Wei}, \citenamefont {Zou}, \citenamefont {Sheng},
  \citenamefont {Chen},\ and\ \citenamefont {Xing}}]{PhysRevB.107.035306}%
  \BibitemOpen
  \bibfield  {author} {\bibinfo {author} {\bibfnamefont {H.}~\bibnamefont
  {Geng}}, \bibinfo {author} {\bibfnamefont {J.~Y.}\ \bibnamefont {Wei}},
  \bibinfo {author} {\bibfnamefont {M.~H.}\ \bibnamefont {Zou}}, \bibinfo
  {author} {\bibfnamefont {L.}~\bibnamefont {Sheng}}, \bibinfo {author}
  {\bibfnamefont {W.}~\bibnamefont {Chen}},\ and\ \bibinfo {author}
  {\bibfnamefont {D.~Y.}\ \bibnamefont {Xing}},\ }\bibfield  {title} {\bibinfo
  {title} {Nonreciprocal charge and spin transport induced by non-hermitian
  skin effect in mesoscopic heterojunctions},\ }\href
  {https://doi.org/10.1103/PhysRevB.107.035306} {\bibfield  {journal} {\bibinfo
   {journal} {Phys. Rev. B}\ }\textbf {\bibinfo {volume} {107}},\ \bibinfo
  {pages} {035306} (\bibinfo {year} {2023})}\BibitemShut {NoStop}%
\bibitem [{\citenamefont {Shao}\ \emph {et~al.}(2024)\citenamefont {Shao},
  \citenamefont {Geng}, \citenamefont {Liu}, \citenamefont {Lado},
  \citenamefont {Chen},\ and\ \citenamefont {Xing}}]{PhysRevLett.132.156301}%
  \BibitemOpen
  \bibfield  {author} {\bibinfo {author} {\bibfnamefont {K.}~\bibnamefont
  {Shao}}, \bibinfo {author} {\bibfnamefont {H.}~\bibnamefont {Geng}}, \bibinfo
  {author} {\bibfnamefont {E.}~\bibnamefont {Liu}}, \bibinfo {author}
  {\bibfnamefont {J.~L.}\ \bibnamefont {Lado}}, \bibinfo {author}
  {\bibfnamefont {W.}~\bibnamefont {Chen}},\ and\ \bibinfo {author}
  {\bibfnamefont {D.~Y.}\ \bibnamefont {Xing}},\ }\bibfield  {title} {\bibinfo
  {title} {Non-hermitian moir\'e valley filter},\ }\href
  {https://doi.org/10.1103/PhysRevLett.132.156301} {\bibfield  {journal}
  {\bibinfo  {journal} {Phys. Rev. Lett.}\ }\textbf {\bibinfo {volume} {132}},\
  \bibinfo {pages} {156301} (\bibinfo {year} {2024})}\BibitemShut {NoStop}%
\end{thebibliography}%

\end{document}